\documentclass[sigconf, nonacm]{acmart}

\newcommand\vldbdoi{XX.XX/XXX.XX}
\newcommand\vldbpages{XXX-XXX}
\newcommand\vldbvolume{14}
\newcommand\vldbissue{1}
\newcommand\vldbyear{2020}

\newcommand\vldbtitle{\shorttitle} 
\newcommand\vldbavailabilityurl{https://github.com/ljqcodelove/ContTune}
\newcommand\vldbpagestyle{empty} 

\usepackage{xcolor}
\usepackage{algorithm}
\usepackage{algpseudocode}
\usepackage{amsmath}
\usepackage{bm}
\usepackage{multirow}
\usepackage{float}
\usepackage[section]{placeins}
\usepackage{subcaption}
\usepackage[normalem]{ulem}

\usepackage{tabularx}

\begin{document}
\title{ContTune: Continuous Tuning by Conservative Bayesian Optimization for Distributed Stream Data Processing Systems}








\author{
  Jinqing Lian{$^{\star}$},~~~Xinyi Zhang{$^{\S}$},~~~Yingxia Shao{$^{\star}$},~~~Zenglin Pu{$^{\star}$},~~~Qingfeng Xiang{$^{\star}$},~~~Yawen Li{$^{\star}$},~~~Bin Cui{$^{\S}$}
}
\affiliation{%
  \institution{
  {$^{\star}$} Beijing University of Posts and Telecommunications \\
  {$^{\S}$}School of CS \& Key Laboratory of High Confidence Software Technologies, Peking University \\        
  }
}
\email{{jinqinglian,shaoyx,pzl_bupt,xiangqingfeng,warmly0716}@bupt.edu.cn, {zhang_xinyi,bin.cui}@pku.edu.cn} 


\begin{abstract}
The past decade has seen rapid growth of distributed stream data processing systems. 
Under these systems, a stream application is realized as a Directed Acyclic Graph (DAG) of operators, where the level of parallelism of each operator has a substantial impact on its overall performance. 
However, finding optimal levels of parallelism remains challenging. 
Most existing methods are heavily coupled with the topological graph of operators, unable to  efficiently tune under-provisioned jobs. 
They either insufficiently use previous tuning experience by treating successively tuning independently, or explore the configuration space aggressively, violating the Service Level Agreements (SLA). 

To address the above problems, we propose ContTune, a continuous tuning system for stream applications. 
It is equipped with a novel Big-small algorithm, in which the Big phase decouples the tuning from the topological graph by decomposing the job tuning problem into sub-problems that can be solved concurrently. 
We propose a conservative Bayesian Optimization (CBO) technique in the Small phase to speed up the tuning process by utilizing the previous observations. 
It leverages the state-of-the-art (SOTA) tuning method as conservative exploration to avoid SLA violations. 
Experimental results show that ContTune reduces up to 60.75\% number of reconfigurations under synthetic workloads and up to 57.5\% number of reconfigurations under real workloads, compared to the SOTA method DS2. 
\end{abstract}

\maketitle

\pagestyle{\vldbpagestyle}
\begingroup\small\noindent\raggedright\textbf{PVLDB Reference Format:}\\
{Jinqing Lian, Xinyi Zhang, Yingxia Shao, Zenglin Pu, Qingfeng Xiang, Yawen Li, Bin Cui}. \vldbtitle. PVLDB, \vldbvolume(\vldbissue): \vldbpages, \vldbyear.\\
\href{https://doi.org/\vldbdoi}{doi:\vldbdoi}
\endgroup
\begingroup
\renewcommand\thefootnote{}\footnote{\noindent
This work is licensed under the Creative Commons BY-NC-ND 4.0 International License. Visit \url{https://creativecommons.org/licenses/by-nc-nd/4.0/} to view a copy of this license. For any use beyond those covered by this license, obtain permission by emailing \href{mailto:info@vldb.org}{info@vldb.org}. Copyright is held by the owner/author(s). Publication rights licensed to the VLDB Endowment. \\
\raggedright Proceedings of the VLDB Endowment, Vol. \vldbvolume, No. \vldbissue\ %
ISSN 2150-8097. \\
\href{https://doi.org/\vldbdoi}{doi:\vldbdoi} \\
}\addtocounter{footnote}{-1}\endgroup

\ifdefempty{\vldbavailabilityurl}{}{
\vspace{.3cm}
\begingroup\small\noindent\raggedright\textbf{PVLDB Artifact Availability:}\\
The source code, data, and/or other artifacts have been made available at \url{\vldbavailabilityurl}.
\endgroup
}

\section{Introduction}

\begin{figure}[t]
  \includegraphics[width=0.475\textwidth]{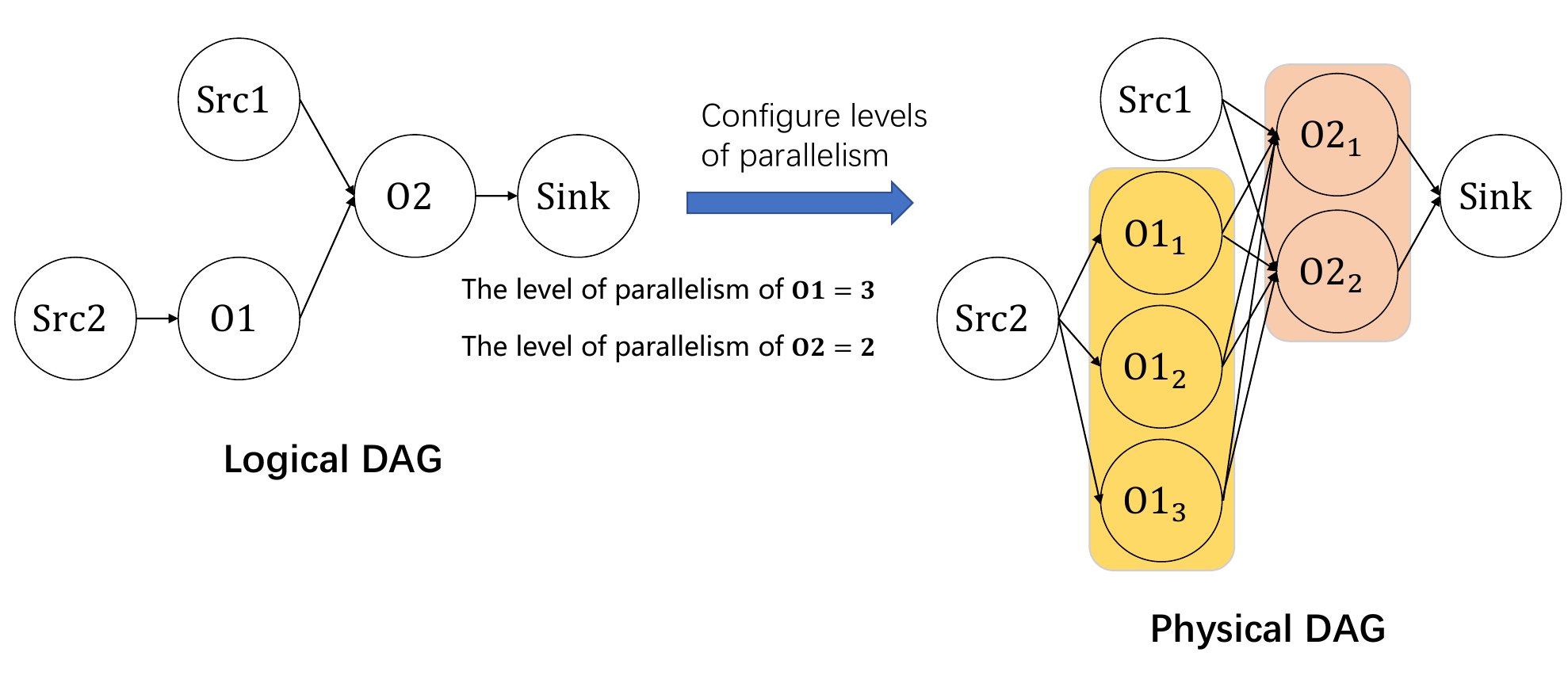}
  \caption{Logical and physical DAG of a stream job.}
  \Description{DAG of Flink jobs.}
  \label{f0}
\end{figure}

In the past decade,  distributed stream data processing systems have been widely used and deployed to handle the big data.
Several mature production systems have emerged, including Flink~\cite{carbone2015apache}, Storm~\cite{toshniwal2014storm}, Spark Streaming~\cite{zaharia2013discretized}, Heron~\cite{kulkarni2015twitter}, and Samza~\cite{noghabi2017samza}.
They can timely analyze the unbounded stream data with low latency and high throughput. 
In these systems, an analytical job is generally abstracted as a Directed Acyclic Graph (DAG) of operators, whose levels of parallelism are configurable. 
The levels of parallelism refer to the configuration of the number of physical instances used by each operator in a job. 
These configurations directly determine the allocation of resources for each operator and have a significant impact on the performance of the job, such as latency and throughput~\cite{cardellini2022runtime,roger2019comprehensive}. 
Figure~\ref{f0} is an example of a job in Apache Flink~\cite{carbone2015apache}, and the level of parallelism of operator O1 is three and the level of parallelism of operator O2 is two.
Therefore, to reduce the Total Cost of Ownership (TCO) while satisfying the Service Level Agreements (SLA), it is critical to set the optimal levels of parallelism.

Given a stream application, configuring the optimal levels of parallelism is non-trivial. 
First, there is no principled way to manually find the optimal levels of parallelism. 
Engineers typically try several configurations and pick the one that satisfies the SLA with minimum resource used~\cite{floratou2017dhalion}. 
Second, considering the dynamic and long-running (i.e., 24/7)  stream data, engineers are required to continuously tune the levels of parallelism so as to adapt to variable workloads. 
As a result, developing effective systems to automatically configure the levels of parallelism has attracted increasing interest from academia and industry~\cite{kalavri2018three,floratou2017dhalion,mei2020turbine,Liu2022OnlineRO,fischer2015machines,mai2018chi,mao2021trisk}. 

Researchers have put considerable efforts into studying parallelism tuning, which can be classified into three categories. 
The first category is rule-based methods~\cite{floratou2017dhalion, ahmad2005distributed,castro2013integrating,gulisano2012streamcloud,gedik2013elastic,heinze2014latency,zaharia2013discretized,xu2016stela}. 
Their tuning policy is usually expressed in simple rules, e.g., if CPU utilization is larger than  70\% then increase the levels of parallelism until CPU utilization is smaller than 70\%. 
The second category is linearity-based methods (e.g., DS2~\cite{kalavri2018three} and Turbine~\cite{mei2020turbine}) which dynamically configure the analytical jobs by linearly increasing or decreasing the levels of 
parallelism. 
The third category is Bayesian Optimization-based methods, represented by Dragster~\cite{Liu2022OnlineRO} and Fischer~\cite{fischer2015machines}. 
They adopt Bayesian Optimization which  utilizes a surrogate model to suggest the promising levels of parallelism and updates the surrogate model based on the effect of the suggested levels of parallelism. 
Owing to these efforts, parallelism tuning can be automated, which largely saves the expensive human-labors.
However, when applying these methods to configure real-world analytical jobs, we encounter several issues from the following perspectives:

\textbf{Inefficiency of tuning under-provisioned jobs.}
The symptom of under-provisioning (e.g., backpressure) usually occurs when the input load increases (workload spikes) and causes SLA violations, which are often associated with significant financial penalties~\cite{Gong2010PRESSPE}.
To configure the under-provisioned job, there are two types of methods: ``find bottleneck and tune it'' and ``workload estimation''. 
The former~\cite{floratou2017dhalion, Liu2022OnlineRO, mao2021trisk, xu2016stela, castro2013integrating, gulisano2012streamcloud} enlarges the level of parallelism of the bottleneck operator one by one~\cite{kalavri2018three}.
Under such an approach, the operator could become a bottleneck repeatedly, influenced by the other tuned operators. 
This is because the operators of a job follow the producer-consumer model, each operator serves as both a producer for its downstream operators and a consumer for its upstream operators in the DAG. 
When scaling up a bottleneck operator, it increases the workloads of its upstream operators as consumers and the workloads of its downstream operators as producers. 
The increased workload can potentially lead to the emergence of new bottleneck operators (as shown in~\cite{Liu2022OnlineRO, floratou2017dhalion, gedik2013elastic}), leading to an increase in the number of reconfigurations. 
So this approach might interrupt the running job frequently and takes a long time to converge to optimal levels of parallelism.  
The latter~\cite{kalavri2018three, mei2020turbine, fischer2015machines, lohrmann2015elastic, fu2017drs} estimates the real upstream data rate and suggests corresponding levels of parallelism to sustain the upstream data.
However, the real upstream data rate cannot be accurately estimated when the job is under-provisioned, specifically for jobs containing \textit{stateful} operators (e.g., join and window operators)~\cite{lombardi2017elastic}.
Besides, the relationship between the configured levels of parallelism and the sustained datas is non-linear and multi-modal. 
They adopt simple approximation (e.g., linear function) and cannot characterize the complex relationship~\cite{Liu2022OnlineRO}.

\textbf{Insufficiency of using previous tuning experience. }
In front of the long-running stream application with inevitable workload variations, most existing methods treat the successively tuning independently, named as the \textit{one-shot parallelism tuning}. 
To be concrete, whenever the parallelism tuning of the analytical jobs is triggered according to the changes of workload, such approaches search for the optimal levels of parallelism from the scratch and do not utilize any observations from previous tuning. 
\textit{One-shot parallelism tuning} is inefficient for the dynamic unbounded stream data and causes a large number of reconfigurations~\footnote{To test a candidate level of parallelism, it requires a reconfiguration which is a time-consuming step. An efficient tuning method finds the optimal level of parallelism with a few (or minimal) number of reconfigurations.}. 
A bad case is that when the job encounters the historical workload (i.e., the workload of stream data has been processed before), the historical optimal level of parallelism can be reused without tuning from the beginning. 
As far as we know, Dragster~\cite{Liu2022OnlineRO} and Turbine~\cite{mei2020turbine} are the only two parallelism tuning methods that utilize the historical information, called \textit{continuous tuning} in this paper. 
Dragster utilizes Bayesian Optimization to find the optimal levels of parallelism for a given upstream data rate. 
However, Dragster tends to aggressively explore the entire configuration space of the levels of parallelism, resulting in frequent violations of the SLA. 
Besides, since Dragster establishes separate Bayesian Optimization models to find the optimal level of parallelism for different upstream data rates of each operator, the reuse of previous tuning experience is only possible when the upstream data rates are identical. 
However, in practice, the range of upstream data rates is wide, making Dragster to rarely reuse the tuning experience.
Turbine makes predictions for future workloads by using historical workloads to determine whether the new configuration is optimal for the predicted future workload, and does not use historical data to accelerate tuning itself. 
In summary, the insufficiency of using the historical tuning experience makes the existing approach inefficient when handling the dynamic workload, and the continuous tuning problem is not well studied yet. 

\textbf{Our approach.}
To address the above challenges, we propose ContTune, a continuous tuning system for elastic stream processing. 
ContTune is equipped with a novel Big-small algorithm, in which the Big phase first decouples the tuning from the topological graph by decomposing the job tuning problem into $N$ sub-problems (discussed in Section~\ref{sec:decom}).
The $N$ sub-problems can be tuned by the Small phase concurrently, largely reducing the number of reconfigurations.
On the basis of the Big-small algorithm, ContTune prioritizes SLA -- it quickly allocates sufficient resources for the under-provisioned jobs in the Big phase and further improves the resource utilization in the Small phase.
Besides, we design a conservative Bayesian Optimization (CBO) technique to speed up the tuning process by utilizing historical observations. 
Compared with vanilla Bayesian Optimization, CBO leverages SOTA \textit{one-shot parallelism tuning} methods~\cite{kalavri2018three, mei2020turbine} as conservative exploration to avoid the SLA violations caused by aggressive exploration. 
Specifically, CBO has two modules: (1) conservative exploration, which utilizes SOTA one-shot parallelism tuning methods to avoid aggressive exploration and warm up the tuning when having insufficient historical observations; 
(2) fast exploitation, which utilizes historical observations to suggests the levels of parallelism according to an acquisition function. 
When compared to Dragster and Turbine, CBO leverages historical observations to establish the relationship between levels of parallelism and the corresponding processing abilities, which is constant and can be used to deal with different upstream data rates.  
On the basis of this relationship, ContTune could quickly find the minimum level of parallelism whose processing ability is lager than the upstream data rate. 
We theoretically prove that ContTune finds optimal levels of parallelism with $O(1)$ average complexity of the number of reconfigurations. 
Specifically, we make the following contributions:

\begin{itemize}
\item {} We propose the Big-small algorithm to tune levels of parallelism for distributed stream data processing systems. The ``Big phase'' can decouple the tuning methods from the topological graph and the ``Small phase'' can concurrently tune all operators. Meanwhile, the Big-small algorithm prioritizes SLA to meet online tuning requirements. 
\item {} We propose CBO to cope with long-running jobs by using historical observations to fit the relationship between the levels of parallelism and processing abilities as \textit{fast exploitation}.
Besides, it first uses \textit{one-shot parallelism tuning} SOTA methods as \textit{conservative exploration} in order to avoid aggressive exploration in vanilla Bayesian Optimization. 
\item {} We implement the proposed method and evaluate on standard benchmarks and real workloads. 
Compared with the SOTA method DS2, ContTune reduced up to \textbf{60.75\%} number of reconfigurations under synthetic workloads and up to \textbf{57.5\%} number of reconfigurations under real workloads. 
\end{itemize}

\section{PRELIMINARY}
We introduce more details of basic concepts such as stream jobs, logical DAG, physical DAG, backpressure, reconfiguration, stateless operators and stateful operators in this section, and formulate the tuning problem. 

\noindent
\subsection{Stream Processing Jobs in DSDPS}
We target at configuring the job (i.e., a stream processing application) in  distributed stream data processing systems (DSDPSs) that are \textit{Data Parallelization}~\cite{roger2019comprehensive}. 
\textit{Data Parallelization} executes one operator on multiple instances. 
The count of these instances is called as the level of parallelism of the operator. 
\textit{Data Parallelization} is commonly supported by DSDPSs, such as Esper~\cite{Esper2019}, Storm~\cite{toshniwal2014storm}, Heron~\cite{kulkarni2015twitter}, Spark Streaming~\cite{zaharia2013discretized}, Flink~\cite{carbone2015apache}, and these systems~\cite{mayer2017minimizing, wu2015chronostream, nasir2015power, nasir2016two, saleh2015partitioning, koliousis2016saber, zacheilas2016dynamic, gedik2016pipelined, mayer2016graphcep, rivetti2016online, schneider2016dynamic, katsipoulakis2017holistic, mayer2017spectre, mencagli2018harnessing, mencagli2018elastic, mencagli2017parallel}. 


\textbf{Logical DAG.} A job (i.e., a stream processing application) in DSDPSs can be modeled as a \textit{logical} Directed Acyclic Graph (DAG) as shown in the left part of Figure~\ref{f0}, denoted as $G = (vertices$ $, edges)$, where the performance of each operator heavily depends on the others, and $vertices$ indicate the operators of the job and $edges$ indicate the passed records (workload) between operators. 
Specifically, operators that only have outgoing edges are \textit{sources}, and operators that only have incoming edges are \textit{sinks}.

\textbf{Physical DAG.} We denote a job running on the given physical instances as a \textit{physical} DAG. 
Figure~\ref{f0} shows a \textit{logical} DAG and its corresponding \textit{physical} DAG for Nexmark Q3~\cite{tucker2008nexmark,Beam,NEXMark} with two sources and one sink. 
Configuring the levels of parallelism of a job decides the number of physical instances for each operator. 
In Figure~\ref{f0}, operators O1 and O2 execute with three and two instances, 
equivalent to their level of parallelism being set three and two.

\textbf{Backpressure. }\label{sec::Backpressure} Backpressure is a mechanism that propagates overload notifications from operators backward to sources so that data emission rates can be throttled to alleviate overload~\cite{cardellini2022runtime}. 
It is a symptom observed in under-provisioned jobs. 
When this happens, workloads that cannot be immediately processed by the sources will not be discarded and are usually kept in the queue~\cite{networkbuffer, lombardi2017elastic}. 

\textbf{Reconfiguration. } The job requires reconfigurations to change the levels of parallelism. 
Each DSDPS enables different reconfigurations methods. Table~\ref{table:wayOfReconfiguration} shows that Flink and Heron need to redeploy (stop and restart), and Trisk adopts a partial pause-and-resume scheme. For all DSDPSs, efficient tuning method finds the optimal level of parallelism using small number of reconfigurations. 

\begin{table}[t]
\centering
\caption{Summary of the reconfiguration methods of existing DSDPSs.}
\label{table:wayOfReconfiguration}
\begin{tabular}{cc}
\hline
Method    & Reconfiguration methods \\ \hline
Flink~\cite{carbone2015apache}     & Redeploy                   \\ \hline
Heron~\cite{kulkarni2015twitter}     & Redeploy                   \\ \hline
Seep~\cite{castro2013integrating}      & Partial redeploy           \\ \hline
Rhino~\cite{del2020rhino}     & Partial update             \\ \hline
Megaphone~\cite{hoffmann2019megaphone} & Non-stop partial update    \\ \hline
Chi~\cite{mai2018chi}       & Partial update            \\ \hline
Trisk~\cite{mao2021trisk}       & Partial update            \\ \hline
\end{tabular}
\end{table}

The operator of a DSDPS job could be stateless or stateful:
(a) \textbf{Stateless Operator.} The data processed by the stateless operator is only relevant to the current operator and the stateless operator does not store the state from previous processing. 
Examples of stateless operators are filter and rescaling. 
(b) \textbf{Stateful Operator.} The data received by the stateful operator will be stored as state information for computation, such as window~\cite{golab2003issues} and join.

\subsection{Problem Definition and Terminology}
We formulate the parallelism tuning problem and discuss the related terminology. 
Table~\ref{Notation} summarizes the notations.

\begin{table}[t]
  \caption{Notations in this paper.}
  \label{Notation}
  \begin{tabular}{ll}
    \toprule
    \textbf{Symbol} & \textbf{Description}\\
    \midrule
    $G$ & logical dataflow Directed Acyclic Graph \\
    $N$ & number of operators in $G$ ($N > 1$) \\
    $\hat{\lambda}$ & aggregated observed upstream data rates\\
    $\lambda$ & real upstream data rates\\
    $T_{u}$ & useful time for an operator\\
    $H^t$ & historical observations with size $t$\\
    $o_{i}$ & the $i^{th}$ operator in $G$\\
    $PA$ & the real processing ability\\
    $p^{j}$ & the level of parallelism at iteration $j$\\
    $p^{max}$ & the max level of parallelism in $H^t$\\
    $p^{now}$ & the now level of parallelism of an operator\\
    $P_{job}$ & the levels of parallelism of operators in $G$ given \\ & by CBO\\
    $GP$ & Gaussian Process model \\
    $\alpha$ & a threshold for scoring function\\
    $d_{nearest}$ & the nearest distance between $p^{now}$ and the \\ &  observed levels of parallelism in $H^t$\\
    $Sg$ & the known region segment\\
    $len_{all}$ & the total length of merged region segments\\
    $\rho$ & $\rho$ tuning times\\
    $\chi$ & CBO uses $\chi$ tuning times in $\rho$ tuning times \\
    $\phi$ & the maximal number of reconfigurations of SOTA \\ & method introduced by CBO of each tuning \\
    $\omega$ & $\omega$ tuning times for fast exploitation to converge  \\
    $W_{u}$ & the workload unit of synthetic workloads\\
  \bottomrule
\end{tabular}
\end{table}

\noindent\textbf{Parallelism Tuning Problem.}
Given a logical DAG of a job with $N$ operators, the source operators generate records at a rate, defined by application data sources (sensors, stock market feeds, etc.)~\cite{kalavri2018three}. 
To maximize system throughput, the physical DAG must sustain the full source rates. 
This means that each operator must be able to process data without backpressure. 
Parallelism tuning aims to find the minimal level of parallelism per operator in the physical DAG that sustains all source rates (i.e., satisfying the SLA). 
Since changing the level of parallelism of the operator requires costly reconfiguration, we additionally want to find the optimal level of parallelism in which each operator can sustain its real upstream data rate via fewer reconfigurations.

\noindent\textbf{Upstream Data Rate.}
An upstream data rate $\hat{\lambda}$ denotes the aggregated number of observed records (i.e., workload) that an operator receives from its all upstream operators per unit of time. 
Given a DAG, the observed upstream data rate is affected by the source rate  and the processing ability of operators in the DAG (following the producer-consumer model). 
When all operators can process their upstream data rate (i.e., no backpressure occurs), the upstream data rate is only affected by the source rate. 
Such an observed upstream data rate is denoted as the real upstream data rate $\lambda$.

\noindent\textbf{Useful Time.}
Useful time $T_u$ is the time that an operator executes in an ideal setting where it never has to wait to obtain input or push output.
It differs from the total observed execution time.
$T_u$ is the total time that an operator spends in serialization, processing and deserialization~\cite{kalavri2018three}.

\noindent\textbf{Processing Ability.} The processing ability  $PA$  denotes how many records an operator can process  per unit of useful time.
We use the same methodology as DS2~\cite{kalavri2018three} to obtain $PA$: 

\begin{equation}\label{equ:1}
PA = \frac{\hat{\lambda}}{T_{u}}.
\end{equation}

An operator's processing ability is affected by its level of parallelism but does not increase linearly  with it~\cite{kalavri2018three, Liu2022OnlineRO}.
After applying a given level of parallelism (denoted as $p_i$) for an operator $o_i$, we could obtain the processing ability of $o_i$  (i.e., $PA(p_i)$)  according to Equation~\ref{equ:1}.
We use $H_i^t$ to denote the historical observations with size $t$ for operator $o_i$ under different levels of parallelism, i.e.,  $H_i^t$= $\left\{\left\langle  p_i^j,PA(p_i^j) \right\rangle \right\}_{j=1}^t$, where $p_i^j$ denotes the level of parallelism for operator $o_i$ at iteration $j$.

\section{System Overview}

\begin{figure}[t]
  \includegraphics[width=0.495\textwidth]{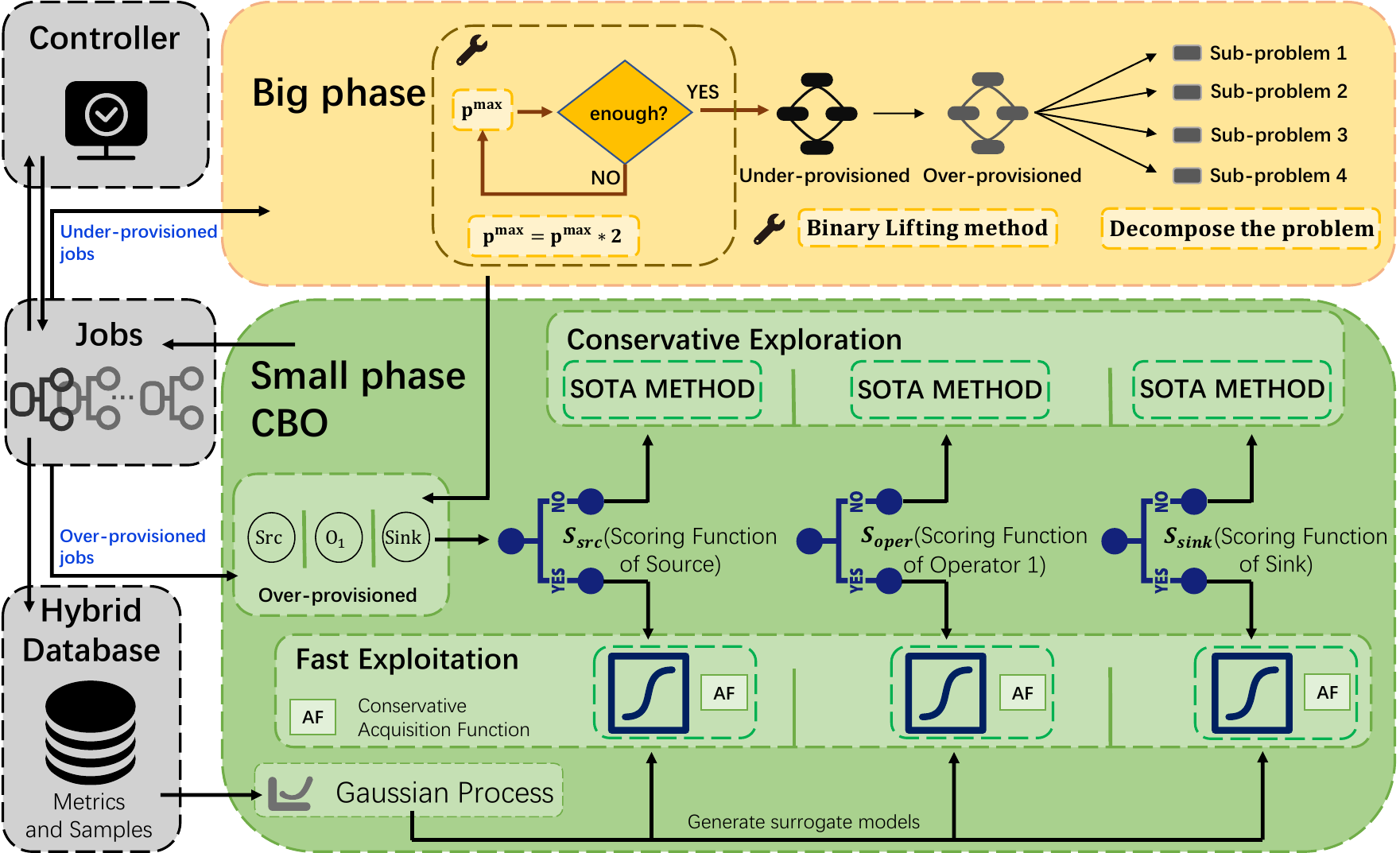}
  \caption{Overall Architecture of ContTune.}
  \Description{The overall architecture of ContTune.}
  \label{f1}
  \vspace{-2em}
\end{figure}

Figure~\ref{f1} presents the overview of ContTune. 
The controller queries the job-generated metrics and determines whether a job needs tuning based on the conditions (discussed in Section~\ref{sec:impl}). 
When a job tuning is triggered, the controller checks the state of the job. 
It detects symptoms of over- or under-provisioning (e.g. backpressure).
Then under-provisioned jobs go through the Big and Small phases, while over-provisioned jobs directly enter the Small phase. 
The Big phase enlarges the levels of parallelism of the job, following the \textit{Binary Lifting} method which quickly eliminates the under-provisioned state. 
The Small phase is executed when the job is not under-provisioned. 
It finds the minimal level of parallelism of each operator that can sustain the real upstream data rate via conservative Bayesian Optimization (CBO).
CBO adopts two strategies to find the optimal levels of parallelism: \textit{fast exploitation} and \textit{conservative exploration}. 
The fast exploitation utilizes historical observations. 
It fits Gaussian Processes (GP) on the observations and suggests the levels of parallelism according to a carefully designed acquisition function that guarantees the SLA. 
The conservative exploration utilizes SOTA \textit{one-shot parallelism tuning} methods to avoid aggressive exploration and warm up the learning of GP. 
ContTune adopts a scoring function to balance the fast exploitation and conservative exploration. 
After CBO finds the optimal level of parallelism of each operator, in order to avoid frequent reconfigurations, the controller confirms whether applying the levels of parallelism output by CBO is necessary given the current levels of parallelism (discussed in Section~\ref{sec:impl}). 
If necessary, the controller reconfigures the job with the levels of parallelism output by CBO, otherwise skips this reconfiguration. 
At the end of each tuning, the observed levels of parallelism of operators and their corresponding processing abilities will be added to $H^t$. 

\section{Big-small Algorithm} 
In this section, we first discuss the decomposition of the parallelism tuning problem which can be efficiently solved. 
Then we present the Big phase to decompose the problem and the Small phase to solve the decomposed problem.

\subsection{Decomposing the Parallelism Tuning }\label{sec:decom}

Given a logical DAG of a job with $N$ operators, parallelism tuning aims to find the minimal level of parallelism per operator that sustains all source rates. 
Sustaining all source rates is equivalent to that every operator can process their real upstream data rates. 
Note that the real upstream data rate reflects the real workload of each operator under the producer-consumer model, which is different from the observed one when the job is under-provisioned. 
The real upstream data rate only can be observed when the job is not under-provisioned. 
Therefore, if we can obtain the real upstream data rates of each operator, the parallelism tuning of a job can be decomposed to the parallelism tunings of each operator. 
Each operator can be concurrently tuned to fulfill their corresponding upstream data rate $\lambda_i$.
And there is no need to tune a single bottleneck operator one by one (requiring several reconfigurations), as "find bottleneck and tune it" used by many existing tuning methods (e.g., Dhalion~\cite{floratou2017dhalion}, Dragster~\cite{Liu2022OnlineRO}, IBM Streams~\cite{gedik2013elastic} and GOVERNOR~\cite{chen2017governor}). 
Formally, we aim to solving the following equation to achieve the minimal number of reconfigurations:

\begin{equation}
\begin{aligned} \label{2}
& \mathop{\arg\min}_{p_1,...,p_N} p_i, and \ p_i \leq maximal \ bound \\
& \mbox{subject to}\
PA(p_i) \geq \lambda_i. \\
\end{aligned}
\end{equation}

Other existing tuning methods (e.g., DS2~\cite{kalavri2018three}) use instrumentation of bottleneck detection tools (e.g., SnailTrail~\cite{DBLP:conf/nsdi/HoffmannLLKDWCR18}) to estimate $\lambda$ via selectivities between operators. 
We find that these tuning methods face two problems. 
First, the instrumentation brings additional overhead, which increases the per-record latency (e.g., 13\% as shown in~\cite{kalavri2018three}). 
Second, the estimated $\lambda$ might be inaccurate since selectivities between stateful operators are inaccurate~\cite{lombardi2017elastic}. 
Federico et. al~\cite{lombardi2017elastic} point out that stateful operators have a large standard deviation of the observed selectivities (discussed in ~\cite{kalavri2018three, tu2006load, lombardi2017elastic}) due to their semantics, and it is inaccurate for ``workload estimation'' method to use observed selectivities at a specific moment to represent the selectivities of these operators.
For example, window, a typical stateful operator, splits the infinite stream into ``buckets'' of finite size, over which DSDPSs can apply computations. 
It may obtain a observed selectivity of zero if no ``buckets'' are computed within the observed time. 
Then the inaccurate selectivity is used to estimate the upstream data rates to the downstream operators.
The inaccuracy is propagated over the topological graph, leading to the non-optimal levels of parallelism of the operators. 
We use the Big-small algorithm to tackle these two problems. 
If the job is under-provisioned at the beginning, the Big phase first make the job not in backpressure state and then uses the observed $\hat{\lambda}$ as the real $\lambda$. 
The over-provisioned job at the beginning directly enters the Small phase. 
Then the parallelism tuning problem can be decomposed into $N$ sub-problems that find the minimal level of parallelism per operator whose processing ability is not less than its real upstream data rate, i.e., $PA(p_i) \geq \lambda_i$. 
Then Big phase decouples the parallelism tuning from the topological graph by decomposing the parallelism tuning to $N$ sub-problems. 
And the Small phase can concurrently tune these $N$ sub-problems.

\begin{algorithm}[t]
  \caption{Big-small Algorithm }
  \label{alg:bigsmall}
  \begin{algorithmic}[1]
    \Require
      A stream job with $N$ operators, the maximal level of parallelism observed $p^{max}$ in all $H^t$ for the job
    \Ensure The levels of parallelism suggested for the given job $P_{job}$
    
    \State {// ``Big'' phase}
    \While {the job is under-provisioned} 
        \State {$Flag$ $\leftarrow$ True}
        \For{$i \leftarrow 1 \ldots N$} \label{line:check1}
          \If {${p^{now}}_i$ $\neq$ $p^{max}$}
            \State {$Flag$ $\leftarrow$ False}
          \EndIf \label{line:check2}
        \EndFor
        \If {$Flag$} 
          \state {$p^{max} \leftarrow 2 * p^{max}$} \label{line:double}
        \EndIf
        \For{$i \leftarrow 1 \ldots N$}
          \State {${p^{now}}_i$ $\leftarrow$ $p^{max}$} \label{line:setmax}
        \EndFor
        \State{$P_{job} \leftarrow \left\{p_i^{now}\right\}_{i=1}^N$ and apply $P_{job}$ via a reconfiguration} \label{line:endOfBigPhase}
    \EndWhile
    \State{}
    \State {// ``Small'' phase}
    \For{$i \leftarrow 1 \ldots N$} 
        \State {${\lambda}_{i} \leftarrow {\hat{\lambda}}_{i} $ \ // The job is not under-provisioned now}
    \EndFor
    \State {Use Algorithm~\ref{CBOAlgorithmCode} to get $P_{job}$}
    
  \end{algorithmic}
\end{algorithm}

\subsection{Big Phase and Small Phase}\label{sec:phase}
Algorithm~\ref{alg:bigsmall} illustrates the main procedures of the Big-small algorithm. 
The algorithm has two phases: \textbf{Big} and \textbf{Small}. 
Under-provisioned jobs go through these two phases, while over-provisioned jobs directly enter the Small phase. 
The Big phase increases the efficiency of tuning by first giving sufficient levels of parallelism so that the job is not in backpressure state. 
For the over-provisioned job, the Small phase aims to quickly find minimal levels of parallelism that the job can sustain all source rates, thereby improving resource utilization. 

\noindent
\textbf{Big Phase.}
The Big phase focuses on the fast elimination of operator's backpressure using the \textit{Binary Lifting} method, which can even out the time complexity with the help of historical observations $H^t$ (discussed in Section~\ref{sec:timeComplexityOfContTune}). 
The Big phase maintains the maximal level of parallelism as $p^{max}$, among all the observations in $H^t$. 
All jobs that are at the end of the Big phase, rather than the end of the Small phase, satisfy their SLA, which means the Big phase prioritizes SLA to meet online tuning requirements. 
Specifically, the Big phase first checks if each operator's current level of parallelism $p^{now}_i$ is equivalent to $p^{max}$ (Line~\ref{line:check1} -- Line~\ref{line:check2}). 
If each $p^{now}_i$ is equivalent to $p^{max}$ and the job is still under-provisioned, it indicates that $p^{max}$ is not enough to sustain the upstream data rate, thus ContTune doubles $p^{max}$ (Line~\ref{line:double}). 
Finally, the Big phase sets the current $p^{now}_i$ to $p^{max}$, $i=1,.., N$ via one reconfiguration (Line~\ref{line:setmax}). 
The above process (Line~\ref{line:check1} -- Line~\ref{line:endOfBigPhase}) is repeated until  the job is not under-provisioned. 

\noindent
\textbf{Small Phase.}
In the Small phase, we use CBO (details are discussed in Section~\ref{sec:cbo}) to find the optimal levels of parallelism for the over-provisioned jobs to improve resource utilization while satisfying $PA(p_i))\geq\lambda_i$.

\section{Conservative Bayesian Optimization}\label{sec:cbo}

We adopt Bayesian Optimization to configure the levels of parallelism to improve CPU utilization.
We present how we adopt the BO to suggest the configuration with minimal resource usage while considering the SLA requirements in Section~\ref{sec:BO4ParallelismTuning}. 
To further avoid the SLA violation, we introduce the conservative Bayesian Optimization (CBO) in Section~\ref{sec:trade-offBetweenConservativeExplorationAndExploitation}, which adopts linearity-based methods to replace the aggressive exploration in the vanilla BO.

\subsection{BO for Parallelism Tuning}\label{sec:BO4ParallelismTuning}

As discussed in Section~\ref{sec:decom}, optimizing the whole DAG can be decomposed to optimizing $N$ sub-problems as Equation~\ref{2}.
As guaranteed by the Big phase, $p^{max}$ is set as the maximal bound in Equation~\ref{2}. 
To find the desired $p_i$, one naive method is to evaluate all possible levels of parallelism, which is prohibitively expensive due to the number of required reconfigurations and the violation of SLA. 
To this end, we adopt Bayesian Optimization (BO) to guide the search for desired $p_i$.

BO is a widely-used optimization framework for the efficient optimization of expensive black-box functions. 
It has two key modules: (1) a \textit{surrogate model} that learns the relationship between configurations and the performances, (2) an \textit{acquisition function} that measures the utility of the given configurations according to the estimation of the surrogate model. 
In contrast to evaluating the expensive black-box function, the acquisition function is cheap to compute and can therefore be thoroughly optimized~\cite{wu2019hyperparameter}.
BO works iteratively: it chooses the next configuration to evaluate by maximizing the acquisition function and  then updates the surrogate model based on the augmented observations.
The main challenge of adopting BO is to set up suitable surrogate model and acquisition function for parallelism tuning.

\textbf{Surrogate Model.}
In our BO method, we adopt Gaussian Process (GP) as the surrogate model. 
GP is a non-parametric model that can adaptively adjusts its complexity to fit the data, which allows GP to capture intricate patterns and adapt to various data distributions. 
Besides, it provides well-calibrated uncertainty estimations and closed-form computability of the predictive distribution~\cite{automl}. 
Other data-intensive techniques, e.g., deep learning may struggle with low data efficiency and interpretability. 
We adopt GP to learn the relationship between the levels of parallelism of the operator $o_{i}$ and its processing abilities, based on $H_i^t$. 
Formally, it fits a probability distribution $p(f|p_i,  H_{i}^t)$ of the target function $PA\left(p_i\right)$ on the observations $H_i^t$. 
With the help of GP, given a level of parallelism $p_i$, we can estimate its processing ability as a Gaussian variable with mean $\mu(p_i)$ and variance $\sigma^2(p_i)$ (indicating the confidence level of the estimation): 

\begin{equation}
\begin{aligned}
    \mu(p_i) &= k_*^T K ^{-1} y, \\
    \sigma^2(p_i) &= k_*\big(p_i, p_i\big) - k_*^T {K}^{-1} k,
\end{aligned}\label{equ:gp}
\end{equation}
where $k$ is the covariance function,  $k_*$ denotes the vector of covariances between $p_i$ and all previous observations, $K$ is the covariance matrix of all previously evaluated configurations and $y$ is the observed function values. 
To this end, we can utilize the confidence level to  obtain the  bound of the estimation: $l(p_i)= \mu(p_i)-\beta\sigma(p_i)$ and $u(p_i)= \mu(p_i)+\beta\sigma(p_i)$,  where the parameter $\beta$ controls the tightness of the confidence bounds~\cite{DBLP:conf/icml/SrinivasKKS10}. 
The true function value of $PA(P_i)$ falls into the interval $[l(p_i), u(p_i)]$ with a high probability. 
\vspace{0.5em}

\noindent\textit{Noise Treatment.}
ContTune uses Top-K technique to cope with noise brought by cluster changes, hardware changes or changes in parameters other than parallelism. 
Since the processing ability of the operator might be disturbed by other environmental factors like network latency~\cite{farhat2021klink}, ContTune models the environmental factor as positive additive noise. 
ContTune invokes Top-K technique at the database level and only choose K recent observations for each operator, and for these K observations, ContTune performs mean-reversion on them to obtain the noise-reduced metrics for tuning (discussed in Section~\ref{sec:impl}). 
Using Top-K and mean-reversion, ContTune is efficient by avoiding additional reconfigurations caused by noise.

\textbf{Acquisition Function.}
The sub-problem is essentially a minimization problem with  an unknown constraint, as shown in Equation~\ref{2}.
The desired acquisition function should guide the finding of desired $p_i$ as soon as possible, and avoid the SLA violation during tuning.
Common acquisition functions such as UCB~\cite{srinivas2009gaussian} and Expected Improvement (EI)~\cite{DBLP:conf/acml/NguyenGR0V17} do not support the constraint conditions. 
Recently, Constrained EI (CEI) is proposed to optimize a black-box function with unknown constraints for optimizing the resource usage in data management systems~\cite{DBLP:conf/uai/GelbartSA14,DBLP:conf/sigmod/ZhangWCJT0Z021}:
\begin{equation}
\label{EIC}
arg\max _{p_i}\left( ( p_i^*-p_i ) \times Pr[f(p_i)\ge\lambda ] \right),
\end{equation}
where $p_i^*$ denotes the minimal feasible level of parallelism and $Pr[f(p_i)\ge\lambda ]$ denotes 
the probability of feasibility.
The probability of feasibility guides the search for feasible level of parallelism, while the reduced level of parallelism, i.e., ($ p_i^*-p_i$) encourages improving resource utilization. 
However, CEI does not consider the constraint safety-critical, and it may suggest infeasible levels of parallelism during tuning (e.g., trying the level of parallelism $p_i$ with large $ p_i^*-p_i$ but small $Pr[f(p_i)\ge\lambda ]$).
Once these levels of parallelism are suggested, additional reconfigurations are required to keep the job from under-provisioned. 
To prioritize the SLA while tuning, we make the safety constraint of CEI more strict and use the following acquisition function: 

\begin{equation}\label{equ:acq-our}
\begin{gathered}
arg\max _{p_i}( p_i^*-p_i)I\left( \mu( p_i ) - \lambda_i \right) 
\end{gathered}
\end{equation},
where $I(x)$ is an indicator function:
\begin{equation}
    I\left( x \right) =\begin{cases}
	1&		if\ x\ge 0,\\
	0&		if\ x<0.\\
\end{cases} \\
\end{equation}
In the acquisition function, $I\left( \mu( p_i ) - \lambda_i \right) $ filters the infeasible levels of parallelism based on GP's estimation.
Thus, the SLA guarantee is considered the first priority while tuning.

\subsection{Trade-off between Conservative Exploration and Fast Exploitation}\label{sec:trade-offBetweenConservativeExplorationAndExploitation}

The above acquisition function filters infeasible levels of parallelism based on GP's estimation. 
However, in the region with few observations (i.e., unknown region), the estimation will yield large uncertainty  (e.g., the cold start case).
Exploring the unknown region is inevitable in vanilla BO since it serves as part of learning for the objective functions.
However, aggressive exploration is unfavorable in parallelism tuning, since  the SLA cannot be guaranteed.

To tackle the problem, we propose to utilize linearity-based tuning methods to warm up the learning of GP and cope with sudden changes in workload. 
The linearity-based methods estimate the processing ability per instance and essentially suggest the levels of parallelism of operators based on the ratio between the upstream data rate and the processing ability. 
Since the relationship between the levels of parallelism and processing abilities is non-linear~\cite{Liu2022OnlineRO}, they cannot converge to the optimal level of parallelism in one step. 
But the linearity-based methods are suitable for warming up the GP and being conservative exploration to avoid aggressive exploration. 
Since the aggressive exploration is avoided, CBO can be used in real online environments. 
We refer to the level of parallelism suggested by linearity-based methods for operator $o_{i}$ as $p_i^{lin}$ and the level of parallelism suggested by the acquisition function as $p_i^{acq}$. 
Concretely, CBO applies $p_i^{lin}$, when GP's estimation has large uncertainty -- in other words, when $p_i^{acq}$ falls in the unknown region. 
Otherwise, CBO applies $p_i^{acq}$. 
DS2 is adopted as the linearity-based method in CBO. 
Intuitively, in CBO, the surrogate model and the acquisition function serve as \textit{fast exploitation}, and the linearity-based method serves as \textit{conservative exploration}.

CBO uses a scoring function to achieve the trade-off between conservative exploration and fast exploitation. 
Given a level of parallelism, GP's estimation will be more accurate when the given level of parallelism is closer to the observed levels of parallelism.
Thus, we use a scoring function to decide whether $p_i^{acq}$ falls in the unknown region by how far $p_i^{acq}$ is from the current observations.
Specifically, we  use $d_{nearest}^i$ to denote the minimal value of the distance between $p_i^{acq}$ and the observed level of parallelism in $H_i^t$.
When $d_{nearest}^i$ is smaller than or equal to a threshold, namely $\alpha$, CBO applies $p_i^{acq}$. 
Otherwise, the linearity-based method is adopted and CBO applies $p_i^{lin}$.
And the augmented observation from the linearity-based method is also added to $H_i^t$, as a training sample for GP, which warms up GP's learning.

\begin{algorithm}[t]
  \caption{CBO algorithm}
  \label{CBOAlgorithmCode}
  \begin{algorithmic}[1]
    \Require A stream job with $N$ operators, real upstream data rates for each operator ${\lambda_{i}}_{i=0}^N$,observations for each operator $H_i^t$, a threshold for scoring function $\alpha$ 
 
    \Ensure The suggested levels of parallelism $P_{job}$ for the given job
      
    \State{Initialize $P_{job}$ as an empty list}
    \For{$i \leftarrow 1 \ldots N$} \label{line-sub}
        \State {Fit $GP_i$ based on $H_i^t$ and obtain $p_i^{acq}$ following Equation~\ref{equ:acq-our}}\label{line-p1}

        \State {$d_{nearest}^i \leftarrow  +\infty$ }
        \For {$(p, PA(p))$ in $H_i^t$}
          \State {$d_{nearest}^i \leftarrow \min\left(d_{nearest}^i, \lvert p^{acq}_i - p\rvert\right)$}
        \EndFor
        \If {${d_{nearest}^i} \leq \alpha$}
          \State {Append $p_i^{acq}$} to $P_{job}$
          \Else
           \State {Obtain $p_i^{lin}$ through linearity-based method}
         \State {Append $p_i^{lin}$ to $P_{job}$}
        \EndIf \label{line-p2}
    \EndFor 
    \State {Apply $P_{job}$ via one reconfiguration}\label{line-apply}
    \For{$i \leftarrow 1 \ldots N$}
      \State{Observe $PA(p_i)$ and append $\left\langle  p_i,PA(p_i) \right\rangle$ to $H_i^t$}\label{line-add}
    \EndFor
\State{\textbf{return} $P_{job}$}
  \end{algorithmic}
\end{algorithm}

Algorithm~\ref{CBOAlgorithmCode} presents one tuning step of CBO formally.
CBO deals with each operator separately (Line~\ref{line-sub}), since the tuning problem is decomposed into $N$ sub-problems as Equation~\ref{2}.
CBO first obtains true $\lambda$ as the job is not in backpressure state.
Then it fits a GP model on $H_i^t$ and obtain $p_i^{acq}$ and decides whether to use $p_i^{acq}$ or $p_i^{lin}$ by the scoring function (Line~\ref{line-p1} - Line~\ref{line-p2}).
After all the sub-problems are solved, CBO applies the suggested levels of parallelism via one reconfiguration (Line~\ref{line-apply}) and saves the corresponding observations (Line~\ref{line-add}).

\subsection{Continuous Tuning via CBO}

Given a stream job, the workload of stream data (i.e., source rate) is dynamic and the applied levels of parallelism can become inappropriate for the source rate, thus parallelism tuning will be triggered accordingly. 
Such a scenario is called \textit{continuous tuning}. 
CBO adopts GP as the surrogate model to fit the relationship between the levels of parallelism and processing abilities. 
Although the source rate changes, the relationship between the levels of parallelism and processing abilities of an operator is constant. 
And the surrogate models (i.e., GPs) in CBO can be reused for continuously tuning a stream application in spite of different source rates. 
Intrinsically, the GPs in CBO enable the speedup of target tuning via historical observations.  
In contrast to \textit{one-shot parallelism tuning} methods, CBO is specifically designed to undergo continuous improvement.
This means that as the number of historical observations increases, the tuning performance of CBO also improves correspondingly.
It employs an acquisition function (Equation~\ref{equ:acq-our}) to identify suitable level of parallelism based on GPs for fast identification of the optimal levels of parallelism.
Faced with the presence of noise in metrics, CBO uses the Top-K technique with mean-reversion to deal with it (discussed in Section~\ref{sec:impl}). 

\begin{figure}[t]
  \includegraphics[width=0.475\textwidth]{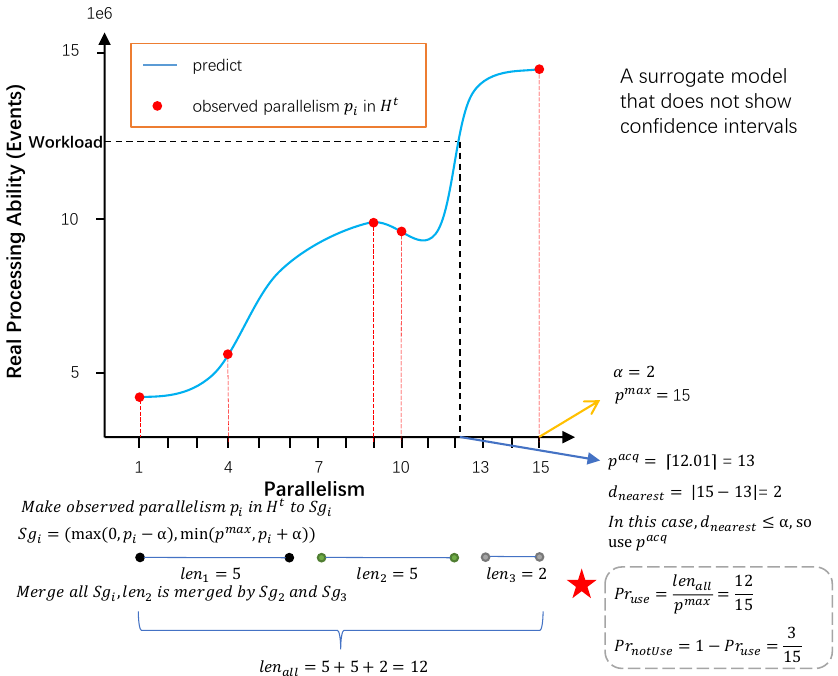}
  \caption{The process of calculating $d_{nearest}$ and probabilities.}\label{fig:plen}
  \Description{The algorithm graph.}
\end{figure}

\section{Efficiency of ContTune}\label{sec:timeComplexity}
We first discuss the convergence of ContTune in Section~\ref{sec:convergenceOfContTune}, 
and analyze its average complexity of the number of reconfigurations in Section~\ref{sec:timeComplexityOfContTune}.

\subsection{Analysis of ContTune Convergence}\label{sec:convergenceOfContTune}
Given an observation $\langle p^j, PA(p^j) \rangle$ in $H^t$, CBO considers the region round $p^j$ with $\alpha$ as the radius to be a known region (i.e., the region with small uncertainty) and the $p^{acq}$ falls in the known region $[\max(p^j - \alpha, 0), \min(p^j + \alpha, p^{max})]$ will be applied. 
The closer the level of parallelism to the observed level of parallelism in $H^t$ the greater the confidence level in the surrogate model, i.e. the smaller the $d_{nearest}$ the greater the confidence level. 
As shown in Figure~\ref{fig:plen}, for $H^t$ with size $t$, therefore there are $t$ known regions, and we refer to the total size for the $t$ regions as $len_{all}$. 
And the maximal bound for the configuration space of the level of parallelism is $p^{max}$. 
Given a random variable ranging from 0 to $p^{max}$, the probability $Pr_{use}$ that it falls in the known region is $\frac{len_{all}}{p^{max}}$. 
As the number of observations increases, the probability that $p^{acq}$ falls in the known region will also increase and the scoring function will be more likely to recommend $p^{acq}$. 
The existence of an upper bound on the workloads implies that the $p^{max}$ is smaller than or equal to a constant value. 
Implying that as tuning proceeds, and the $len_{all}$ is increasing, the probability $\frac{len_{all}}{p^{max}}$ is increasing. 
And CBO converges to fast exploitation over time. 
In fact, the real-world workload is not uniformly distributed, and the probability of hitting the fast exploitation is not less than $\frac{len_{all}}{p^{max}}$. 

Figure~\ref{fig:plen} presents a concrete example. 
For an operator, CBO has its five levels of parallelism (1,4,9,10,15) in $H^t$ with corresponding processing abilities ($PA(1), PA(4), PA(9), PA(10), PA(15)$), and the surrogate model fitted by $H^t$ of this operator is shown in Figure~\ref{fig:plen} without showing the confidence interval. 
In this tuning, the \textit{workload} indicates the upstream data rates received by this operator. 
There are five known regions, and $len_{all} = 12$. 
Because $p^{max} = 15$, we get $Pr_{use} = \frac{12}{15} = \frac{4}{5}$ based on this $H^t$. 
CBO recommends $p^{acq} = 13$ at this time, and the nearest observed level of parallelism in $H^t$ from $p^{acq}$ is 15. The $d_{nearest}$ is 2, and CBO sets $\alpha$ to 2, and $d_{nearest} \leq \alpha$, so CBO recommends $p^{acq}$ in this round of tuning, otherwise, CBO recommends $p^{lin}$. 

\subsection{Average Complexity of the Number of Reconfigurations of ContTune}\label{sec:timeComplexityOfContTune}

In the scenario of one-shot parallelism tuning, the efficiency of a tuning method is often evaluated based on the convergence speed. 
However, considering the long-running nature of jobs in distributed stream data processing systems, it is inappropriate to assess the efficiency of tuning solely based on the speed of convergence in a single tuning time. 
Instead, we use the average complexity of the number of reconfigurations across continuous tuning scenarios as a metric to evaluate the efficiency of ContTune. 
When ContTune tunes long-running jobs, the exploration may delay convergence at a particular tuning time, but it can increase the confidence of the model and yield better results in subsequent tuning times. 

We denote the number of tuning for a job as $\rho$, and the Big phase uses the \textit{Binary Lifting} method, so the complexity of the number of reconfigurations of getting $p^{max}$ (Line~\ref{line:double} in Algorithm~\ref{alg:bigsmall}) is $\log_{2}{p^{max}}$, and the worst-case is that job is under-provisioned at the beginning of each tuning, and needs to reconfigure once (Line~\ref{line:setmax} in Algorithm~\ref{alg:bigsmall}) at each tuning making job not in backpressure state, and during $\rho$ tuning times, the job needs to be reconfigured $\rho$ times in worst. 
So the worst-case number of reconfigurations of the Big phase is $\log_{2}{p^{max}} + \rho$, and the worst average complexity of the number of reconfigurations of the Big phase is:

\begin{equation}\label{time:bigphase}
\begin{gathered}
O\left(\frac{\log_{2}{p^{max}} + \rho}{\rho}\right).
\end{gathered}
\end{equation}

We denote the number of tuning using conservative exploration as $\chi$ $\left(\chi \leq \rho\right)$, then the remaining $\left(\rho - \chi\right)$ number of tuning is used for fast exploitation. 
We denote the maximal number of reconfigurations used for tuning of the SOTA method for each tuning as $\phi$~\footnote{For DS2, $\phi = 3$~\cite{kalavri2018three}.}, therefore conservative exploration introduces $\left(\chi \times \phi\right)$ number of reconfigurations. 
If CBO employs fast exploitation, CBO uses only one reconfiguration in fast exploitation for each tuning in the best case, a simple example is CBO has every level of parallelism ranging from 1 to $p^{max}$ in $H^t$. 
For this best case, fast exploitation introduces $\left(\rho - \chi\right)$ number of reconfigurations. 

In the worst case, if CBO employs fast exploitation when fast exploitation has not yet converged, CBO may find an inappropriate level of parallelism that makes the operator bottlenecked.
The suggested inappropriate level of parallelism found by fast exploitation doesn't belong to $H^t$, because the processing ability in $H^t$ is accurate and not estimated by GP. 
The worst case would employ CBO~\footnote{Using CBO rather than ContTune because we have got the real upstream data rate $\lambda$, so for under-provisioned jobs, the Big phase is not used.} once again. 
We denote the number of reconfigurations introduced by the worst case as $\omega$, and $\omega \leq p^{max}$. 
For this worst case, fast exploitation introduces $\left(\rho - \chi\right) + \omega$ number of reconfigurations. 

So the worst average complexity of the number of reconfigurations of CBO is:

\begin{equation}\label{time:smallphase}
\begin{gathered}
O\left(\frac{\left(\chi \times \phi\right) + \left(\rho - \chi\right) + \omega}{\rho}\right).
\end{gathered}
\end{equation}

The worst average complexity of the number of reconfigurations of ContTune including the Big phase and CBO is:

\begin{equation}\label{time:ContTune}
\begin{gathered}
O\left(\frac{\left(\log_{2}{p^{max}} + \rho\right) + \left(\left(\chi \times \phi\right) + \left(\rho - \chi\right) + \omega\right)}{\rho}\right).
\end{gathered}
\end{equation}
We assume within $\rho$ tuning times, there is an upper bound on the upstream data rates, and correspondingly, there is an upper bound on the $p^{max}$, $\phi \leq 3$ and $\omega \leq p^{max}$, so average complexity of the number of reconfigurations of ContTune is $O(1)$.

\section{Implementation}\label{sec:impl}

\noindent
\textbf{Controller and Conditions.}
The controller determines whether the job is \textit{over-provisioned} or \textit{under-provisioned} based on the metrics reported by the job. 
The controller performs tuning only when the job is over-provisioned or under-provisioned. 
The following describes the definition of over-provisioned and under-provisioned. 

Here are some metrics~\cite{FlinkMetrics} used by the controller:
\begin{itemize}
    \item \textit{backPressuredTimeMsPerSecond}: the time (in milliseconds) this task is in backpressure state per second. 
    \item \textit{idleTimeMsPerSecond}: the time (in milliseconds) this task is idle (has no data to process) per second. 
    \item \textit{busyTimeMsPerSecond}: the time (in milliseconds) this task is busy (neither idle nor in backpressure state) per second.  
    \item \textit{System.CPU.Usage}: overall \% of CPU usage on the machine. 
\end{itemize}

We define two new metrics from the metrics above: 
\begin{itemize}
    \item \textit{allTime}: backPressuredTimeMsPerSecond $+$ idleTimeMsPerSecond $+$ busyTimeMsPerSecond. 
    \item \textit{backPressurePer}: $\frac{backPressuredTimeMsPerSecond}{allTime}$ $\times$ 100\%. 
\end{itemize}

We define two possible states for a job at runtime by comparing its backPressure Percentage \textit{backPressurePer} to the backpressure threshold \textit{backpressureThr}: 
\begin{itemize}
    \item \textit{backpressure}: backPressurePer $\geq$ backpressureThr. 
    \item \textit{non-backpressure}: backPressurePer $<$ backpressureThr.
\end{itemize}

Similarly, we define three possible states for a job at runtime by comparing its CPU core usage to two thresholds (coreMinThr $<$ coreMaxThr): 
\begin{itemize}
    \item \textit{cpuLow}: System.CPU.Usage $<$ coreMinThr. 
    \item \textit{cpuNormal}: coreMinThr $\leq$ System.CPU.Usage $\leq$  coreMaxThr.
    \item \textit{cpuStress}: System.CPU.Usage $>$ coreMaxThr.
\end{itemize}

After defining these states, we then define \textit{over-provisioned} and \textit{under-provisioned} of the job: 
\begin{itemize}
    \item \textit{over-provisioned}: the job is in the \textit{cpuLow} state and the job is in the \textit{non-backpressure} state. 
    \item \textit{under-provisioned}: the job is in the \textit{backpressure} state and the job is in the \textit{cpuStress} state. 
\end{itemize}

\textit{backpressureThr}, \textit{coreMinThr} and \textit{coreMaxThr} are preset thresholds. 
The threshold \textit{coreMaxThr} is not set too high to avoid that CPU utilization being too high and cooling cannot cope with it, and thus makes CPU frequency reduction. 

Faced with the problem of data skew, which is caused by limited memory but not the limited level of parallelism, the existing methods will simply crank up the levels of parallelism due to the detection of backpressure, and then request too many resources, making the tuning crash. 
The controller takes into account both the \textit{backpressure} state and CPU usage state when setting the \textit{under-provisioned} state, preventing the backpressure caused by the lack of other resources from making the tuning method increase CPU resources, which has better robustness. 

Facing a scenario where the workload changes in a small range may cause the controller to repeatedly tune the level of parallelism with a small magnitude. 
Therefore, the controller compares the levels of parallelism $P_{job}$ given by this tuning with the levels of parallelism $P_{job}^{'}$ applied by the job at this time. 
when the following two states are hit: 
\begin{itemize}
    \item $P_{job}$ $<$ $P_{job}^{'}$ and $\frac{1.0 \times P_{job}^{'}}{P_{job}} \geq (1 + decisionThr)$
    \item $P_{job}$ $>$ $P_{job}^{'}$ and $\frac{1.0 \times P_{job}}{P_{job}^{'}} \geq (1 + decisionThr)$
\end{itemize}
, the controller uses the levels of parallelism to redeploy the job, otherwise, it skips this tuning. 
By the way, the sensitivity of tuning is also determined by \textit{decisionThr}, and users can set unique \textit{decisionThr} values for their jobs or adopt the default value in the DSDPS.

\noindent
\textbf{Hybrid Databases.} In terms of database, we adopted the model of hybrid architecture deployment. 
The time series database \textit{Hermes} \footnote{A database used internally by Tencent, similar to ClickHouse~\cite{ClickHouseWebsite}.} is used to store the metrics reported by Flink, and Hermes creates hourly tables, then the time spent on querying current metrics is greatly reduced through time partitioning. 
Since the metrics reported by Flink have many dimensions, we additionally divide the Flink metrics into four hourly tables according to dimensions: \textit{job} table, \textit{node} table, \textit{operator} table and \textit{task} table. 
The tuning-related metrics are stored in the \textit{operator} table. 
By data partitioning, we reduce the amount of data in the table used for queries and reduce the query time. 

The historical observations $H^t$ that are required to establish the surrogate model are stored in MySQL~\cite{MysqlWebsite}. 
The maximum level of parallelism of many jobs is usually $\leq 100$, which means that for these jobs, $H^t$ only needs to store at most (\textit{the number of these jobs} $\times$ \textit{the average number of operators per job} $\times$ \textit{the average maximum level of parallelism per job} $\times$ \textit{K}) rows. 
Due to the limited rows number of $H^t$, insertion and deletion can be directly performed to match the requirement of selecting the nearest (in the time dimension) $K$ observed levels of parallelism and processing abilities in MySQL. 
For a level of parallelism $p$ of an operator, the $K$ \footnote{If the number of records is smaller than $K$, the actual number of records will be used.} processing abilities stored in the $H^t$ are summed and the result of dividing by $K$, and the newly processed data $\left\langle  p,PA(p) \right\rangle$ is used as the processing ability of this level of parallelism used for tuning. 
The level of parallelism $p$ of the ${K+1}^{th}$ observation $\left\langle  p,PA(p) \right\rangle$ from the current one will be deleted from MySQL. 
\section{Experimental Evaluation}

In this section, we evaluate ContTune through end-to-end, dynamic scaling experiments with Flink. 
We verify the efficiency of ContTune in tuning under-provisioned jobs, over-provisioned jobs, stateless jobs and stateful jobs in two scenarios: synthetic workloads and real workloads in Section~\ref{sec:experiment4SyntheticWorkloads} and ~\ref{sec:experiment4RealWorkloads}. 
We then validate the design of ContTune by comparing different acquisition functions and discuss the ablation study of Top-K and mean-reversion in Section~\ref{sec:experimentAnalysisOfContTune}. 

\subsection{Setup}

\noindent
\textbf{Configurations.} 
We run all experiments and use Apache Flink 1.13 configured with 45 TaskManagers, each with 2 slots (maximal level of parallelism per operator = 90) on up to three machines, each with 16 AMD EPYC 7K62 48-Core Processor @2.60GHz cores and 32GB of RAM, running tLinux 2.2 (based on CentOS 7.2.1511).

\noindent
\textbf{Queries.}
We use 6 applications, \textbf{WordCount} chosen from original Dhalion publication benchmark~\cite{floratou2017dhalion} and Queries \textbf{Q1-3, 5, 8} from Nexmark benchmark (multiple queries over a three entities model representing on online auction system)~\cite{Beam, NEXMark, tucker2008nexmark, kalavri2018three}, and 3 real applications \textbf{Video streaming}, \textbf{ETL} and \textbf{Monitoring}. 
\begin{itemize}
    \item \textbf{WordCount}, \textbf{Q1} and \textbf{Q2} contain only stateless operators, such as map and filter and there are 3 operators in WordCount, 3 operators in Q1 and 3 operators in Q2. 
    \item \textbf{Q3} contains incremental join, a stateful record-at-a-time two-input operator and there are 5 operators in Q3. 
    \item \textbf{Q5} and \textbf{Q8} contain two window operators: sliding window, tumbling window join and there are 3 operators in Q5, 4 operators in Q8. 
    \item \textbf{Video streaming} contains 3 operators with huge data for Tencent Meeting. 
    \item \textbf{ETL} contains 8 operators with complex DAG for Wechat. 
    \item \textbf{Monitoring} contains 9 operators with 3 sources and 3 sinks. 
\end{itemize}

\begin{table}[t]
\caption{Workload Unit rate (records/s) configuration for WordCount and Nexmark queries on Apache Flink.}
\label{BaseSource}
\begin{tabular}{c|ccc}
\hline
          & Source &          &         \\ \hline
WordCount & 100K   & -        & -       \\ \hline
          & Bids   & Auctions & Persons \\ \hline
Q1        & 700K   & -        & -       \\ \hline
Q2        & 900K   & -        & -       \\ \hline
Q3        & -      & 200K     & 40K     \\ \hline
Q5        & 80K    & -        & -       \\ \hline
Q8        & -      & 100K     & 60K     \\ \hline
\end{tabular}
\end{table}

\noindent
\textbf{Dynamic Workloads Construction.}
We simulate real-world stream applications by constructing dynamic workloads (i.e., varying their source rate along time). 
We use the workload unit in Table~\ref{BaseSource} and simulate the fluctuation using the full permutation of length 10. 
For example, we generate a period of workloads by varying the source rate as [9$W_{u}$, 2$W_{u}$, 3$W_{u}$, 10$W_{u}$, 1$W_{u}$, 4$W_{u}$, 5$W_{u}$, 8$W_{u}$, 6$W_{u}$, 7$W_{u}$], which  has 10 tuning times. 
To simulate the periodicity, we replicate the 10 different source rate, forming a permutation of 20 source rates, which has 20 tuning times. 
We sample 6 permutations ($per1-6$) for each application, i.e., a total of 120 tuning times for each application. 
According to the mechanism of applied tuning method, each tuning time may bring different reconfigurations, even zero due to that the tuner is not triggered. 

\begin{figure}[t]
  \includegraphics[width=0.475\textwidth]{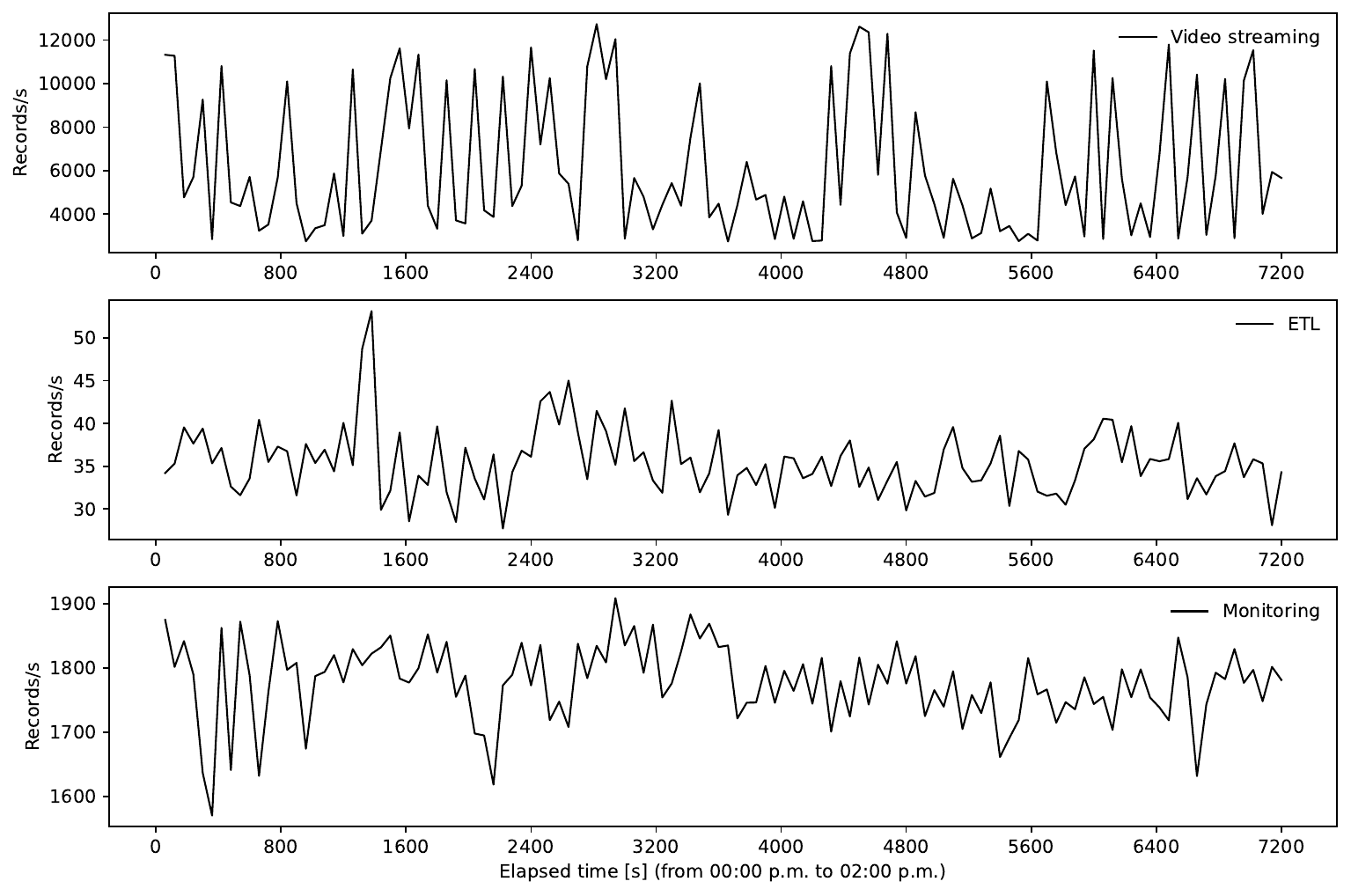}
  \caption{The job workload from Tencent's real cluster from 00:00 p.m. to 02:00 p.m..}
  \Description{The workload from Tencent's real cluster.}
  \label{f4}
\end{figure}

For applications \textbf{Video streaming}, \textbf{ETL} and \textbf{Monitoring}, we collected their real aggregated workloads on Sources from zero p.m. to two p.m. as shown in Figure~\ref{f4}. 

\noindent
\textbf{Baselines.} The baselines are presented below. 
\begin{itemize}
    \item \textbf{Dhalion}~\cite{floratou2017dhalion}: Dhalion is a rule-based method which increases the level of parallelism of an operator suffering from backpressure. 
    We adopted the same rule as in its paper. 
    \item \textbf{DS2}~\cite{kalavri2018three}: DS2 is a linearity-based method and the SOTA parallelism tuning method. 
    We used the same parameters as in its paper. 
    \item \textbf{Big + DS2}: The Big phase first ensures that the job is not in backpressure state and get the real upstream data rate $\lambda$, and then DS2 tunes the backpressure-free job. 
    \item \textbf{Dragster}~\cite{Liu2022OnlineRO}: Dragster is a Bayesian Optimization-based method, which needs to preset the upper bound of the level of parallelism. 
    Dragster provides two tuning methods, ``Online Saddle Point Algorithm'' and a ``Two-level Online Optimization Framework''. 
    The former has shown to accomplish the tuning with fewer reconfigurations, so we used Dragster with ``Online Saddle Point Algorithm''. 
    For the maximal bound, we use $p^{max}$ in ContTune ($\alpha = 3$) as the maximal bound for each query. 
    Specially, Dragster caches hyper parameters of each tuning for the case that the workload has been processed. 
    \item \textbf{ContTune ($\alpha = 0$)}: It uses ContTune to tune the levels of parallelism and sets $\alpha$ to 0. 
    Therefore, it will only apply the observed levels of parallelism in $H^t$ or the levels of parallelism suggested by linearity-based tuning methods.
    Then ContTune ($\alpha = 0$) can be considered as linearity-based tuning methods with cache. 
    \item \textbf{ContTune ($\alpha = 3$)}: It uses ContTune to tune the levels of parallelism and sets $\alpha$ to 3. 
    \item \textbf{Random Search (RS)}: It randomly suggests the levels of parallelism with a given maximal bound.  
    The maximal bound is obtained in the same way as Dragster. 
    The search ends once it finds the same optimal levels of parallelism given by the above methods. 
    Due to the excessive number of reconfigurations required, we enumerate the levels of parallelism and the corresponding processing abilities of all operators beforehand and simulate the search with a program instead (the simulation phase is not accompanied by a real reconfiguration).
\end{itemize}

\subsection{Evaluations on Synthetic Workloads}\label{sec:experiment4SyntheticWorkloads}

We compare ContTune with the baselines on synthetic workloads and make the following observations.

\begin{table*}[t]
    \caption{Evaluations on synthetic workloads. Random Search is simulated due to the large number of reconfigurations. Due to being simulated, Random Search is unable to obtain the real running time, tuning time, and CPU usage. \uwave{A} means the second-best result.} 
    \label{table:syntheticExperiment}
    \begin{subtable}{.5\linewidth}
      \centering
        \caption{Average number of reconfigurations per tuning.}
        \label{table:ResultsUnderContinuous20SimulatedVariableWorkloads}
        \resizebox{1.03\textwidth}{!}{
        \begin{tabular}{c|ccccccc}
        \hline
        Baseline         & WordCount     & Q1            & Q2            & Q3            & Q5            & Q8            & SUM           \\ \hline
        Dhalion          & 3.08          & 5.36          & 4.97          & 3.84          & 5.59          & 3.61          & 4.41          \\ \hline
        DS2              & 1.78          & 2.29          & 2.29          & 1.49          & 3.34          & 3.21          & 2.40          \\ \hline
        Big + DS2        & 1.73          & 2.22          & 2.29          & 1.66           & 3.34          & 3.02         & 2.37  \\ \hline
        Dragster         & 2.75          & 3.85          & 3.85          & 2.75          & 4.95          & 3.85          & 3.67          \\ \hline
        ContTune ($\alpha = 0$) & 1.33          & 1.61          & 1.62          & 1.26          & 2.01          & 1.46          & 1.55          \\ \hline
        ContTune ($\alpha = 3$) & \textbf{1.16} & \textbf{1.32} & \textbf{1.28} & \textbf{1.18} & \textbf{1.55} & \textbf{1.26} & \textbf{1.29} \\ \hline
        Random Search    & 11.16         & 22.72         & 17.43         & 16.18         & 13.36         & 8.72          & 14.93         \\ \hline
        \end{tabular}
        }
    \end{subtable}%
    \begin{subtable}{.5\linewidth}
      \centering
        \caption{Maximal number of requested CPU Cores.}
        \label{table:MaxParallelismResultsUnderContinuous20SimulatedVariableWorkloads}
        \resizebox{0.805\textwidth}{!}{
        \begin{tabular}{c|ccccccc}
        \hline
        Baseline         & WordCount     & Q1            & Q2            & Q3            & Q5            & Q8         \\ \hline
        Dhalion          & 17          & 30         & 29         & 22         & 27         & 13 \\ \hline
        DS2              & \textbf{14}          & \textbf{26}         & \textbf{24}         & \textbf{18}         & \textbf{25}         & \textbf{11}  \\ \hline
        Big + DS2        & 16          & 32         & 32         & 32         & 32         & 16 \\ \hline
        Dragster         & 16          & 32         & 32         & 32         & 32         & 16  \\ \hline
        ContTune ($\alpha = 0$) & 16          & 32          & 32         & 32         &  32         & 16  \\ \hline
        ContTune ($\alpha = 3$) & 16          & 32          & 32         & 32         & 32          & 16 \\ \hline
        Random Search    & 16         & 32         & 32         & 32         & 32         & 16           \\ \hline
        \end{tabular}
        }
    \end{subtable} 
    \begin{subtable}{.5\linewidth}
      \centering
        \caption{End-to-end running time (s).}
        \label{table:E2ERunningTimeResultsUnderContinuous20SimulatedVariableWorkloads}
        \resizebox{1.03\textwidth}{!}{
        \begin{tabular}{c|cccccc}
        \hline
        Baseline         & WordCount     & Q1            & Q2            & Q3            & Q5            & Q8            \\ \hline
        Dhalion          & 81503.71           & 87407.70          & 83633.13          & 82783.87          & 88572.63          & 80546.21         \\ \hline
        DS2              & 76658.36          & 76397.77         & 76144.72          & 75148.37           & 80467.74          & 79644.56            \\ \hline
        Big + DS2        & 76762.78          & 76236.57          & 76248.81          & 75607.61           & 80230.45          & 78697.58           \\ \hline
        Dragster         & 81062.96          & 80691.36          & 80405.64          & 78828.27          & 84426.93          & 81806.26          \\ \hline
        ContTune ($\alpha = 0$) & \uwave{75518.32}\par           & \uwave{75050.01}\par           & \uwave{74931.15}\par          & \uwave{74636.20}\par          &  \uwave{77506.44}\par          & \uwave{75126.16}\par           \\ \hline
        ContTune ($\alpha = 3$) & \textbf{75213.18} & \textbf{74560.88 } & \textbf{74350.65} & \textbf{74439.40} & \textbf{76626.16} & \textbf{74772.08} \\ \hline
        \end{tabular}
        }
    \end{subtable}%
    \begin{subtable}{.5\linewidth}
      \centering
        \caption{Tuning time (s).}
        \label{table:E2ETuningTimeResultsUnderContinuous20SimulatedVariableWorkloads}
        \resizebox{0.945\textwidth}{!}{
        \begin{tabular}{c|cccccc}
        \hline
        Baseline         & WordCount     & Q1            & Q2            & Q3            & Q5            & Q8            \\ \hline
        Dhalion          & 8226.54           & 9441.17          & 8888.33          & 7870.00          & 11928.58          & 8032.83         \\ \hline
        DS2              & 4624.70          & 3991.42         & 4014.15          & 2955.07           & 7128.68          & 7163.79            \\ \hline
        Big + DS2        & 4641.78          & 3901.92          & 4100.27          & 3414.64           & 6941.12          & 6697.58           \\ \hline
        Dragster         & 8569.01          & 7685.27          & 7681.32          & 6757.03          & 11533.06          & 9570.16          \\ \hline
        ContTune ($\alpha = 0$) & \uwave{3394.57}\par           & \uwave{2725.56}\par           & \uwave{2818.33}\par          & \uwave{2476.41}\par          &  \uwave{4209.26}\par          & \uwave{3090.27}\par           \\ \hline
        ContTune ($\alpha = 3$) & \textbf{2947.20} & \textbf{2293.70} & \textbf{2206.18} & \textbf{2304.35} & \textbf{3324.88} & \textbf{2736.19} \\ \hline
        \end{tabular}
        }
    \end{subtable}
    \begin{subtable}{.5\linewidth}
      \centering
        \caption{The percentage of backlogged data.}
        \label{table:BckloggedDataPercentResultsUnderContinuous20SimulatedVariableWorkloads}
        \resizebox{0.94\textwidth}{!}{
        \begin{tabular}{c|ccccccc}
        \hline
        Baseline         & WordCount     & Q1            & Q2            & Q3            & Q5            & Q8         \\ \hline
        Dhalion          & 12.79 (\%)         & 28.31 (\%)         & 19.39 (\%)         & 22.03 (\%)         & 27.43 (\%)         & 7.38 (\%) \\ \hline
        DS2              & 1.92 (\%)        & 4.60 (\%)        & 2.86 (\%)         & 3.53 (\%)          & 8.03 (\%)         & 1.52 (\%)  \\ \hline
        Big + DS2        & 1.52 (\%)        & 3.22 (\%)         & 1.96 (\%)         & 2.33 (\%)          & 7.00 (\%)         & 1.40 (\%) \\ \hline
        Dragster         & 6.08 (\%)        & 7.97 (\%)         & 7.11 (\%)         & 2.52 (\%)         & 7.49 (\%)         & 5.14 (\%)  \\ \hline
        ContTune ($\alpha = 0$) & \textbf{1.01 (\%)}          & \uwave{2.04 (\%)}\par          & \textbf{1.31 (\%)}         & \uwave{1.76 (\%)}\par         &  \uwave{6.86 (\%)}\par         & \uwave{0.66 (\%)}\par  \\ \hline
        ContTune ($\alpha = 3$) & \uwave{1.04 (\%)}\par & \textbf{1.72 (\%)} & \uwave{1.34 (\%)} & \textbf{1.59 (\%)} & \textbf{6.48 (\%)} & \textbf{0.63 (\%)} \\ \hline
        \end{tabular}
        }
    \end{subtable}%
    \begin{subtable}{.5\linewidth}
      \centering
        \caption{Total CPU cost (Cores $\times$ second).}
        \label{table:TotalCostResultsUnderContinuous20SimulatedVariableWorkloads}
        \resizebox{1.05\textwidth}{!}{
        \begin{tabular}{c|ccccccc}
        \hline
        Baseline         & WordCount     & Q1            & Q2            & Q3            & Q5            & Q8         \\ \hline
        Dhalion          & \uwave{540840.16}\par          & \textbf{874697.97}         & \textbf{796300.50}         & \textbf{692169.32}         & \textbf{863432.32}         & \textbf{520995.33} \\ \hline
        DS2              & 550029.28         & 906065.83        & 821041.00         & \uwave{698066.74}\par          & \uwave{876788.87}\par         & \uwave{524793.87}\par  \\ \hline
        Big + DS2        & 562825.71         & 929565.85         & 853817.58         & 732693.26          & 899933.87         & 535530.00 \\ \hline
        Dragster         & 570678.66         & 947174.81         & 865663.73         & 736326.66         & 930962.75         & 543605.93  \\ \hline
        ContTune ($\alpha = 0$) & 548514.47          & \uwave{903635.23}\par          & 820388.86         & 698339.79         &  878424.08         & 533789.47  \\ \hline
        ContTune ($\alpha = 3$) & \textbf{532022.10} & 905940.99 & \uwave{817178.90}\par & 699401.10 & 878743.96 & 533849.47 \\ \hline
        \end{tabular}
        }
    \end{subtable}
\end{table*}

\begin{table}[t]
\caption{CPU cores requested of ContTune and DS2 when they second face the same maximal upstream data rate $10 \times W_{u}$.}
\label{table:CpuCoresWhenMaxThroughput}
\begin{tabular}{c|c|c}
\hline
Queries & ContTune ($\alpha = 3$) & DS2 \\ \hline
WordCount & \textbf{13} CPU cores            & 14 CPU cores \\ \hline
Q1 & \textbf{25} CPU cores           & 26 CPU cores \\ \hline
Q2 & \textbf{23} CPU cores            & 24 CPU cores \\ \hline
Q3 & 18 CPU cores           & 18 CPU cores \\ \hline
Q5 & \textbf{22} CPU cores           & 25 CPU cores \\ \hline
Q8 & \textbf{10} CPU cores           & 11 CPU cores \\ \hline
\end{tabular}
\end{table}

\begin{figure}[t]
  \includegraphics[width=0.475\textwidth]{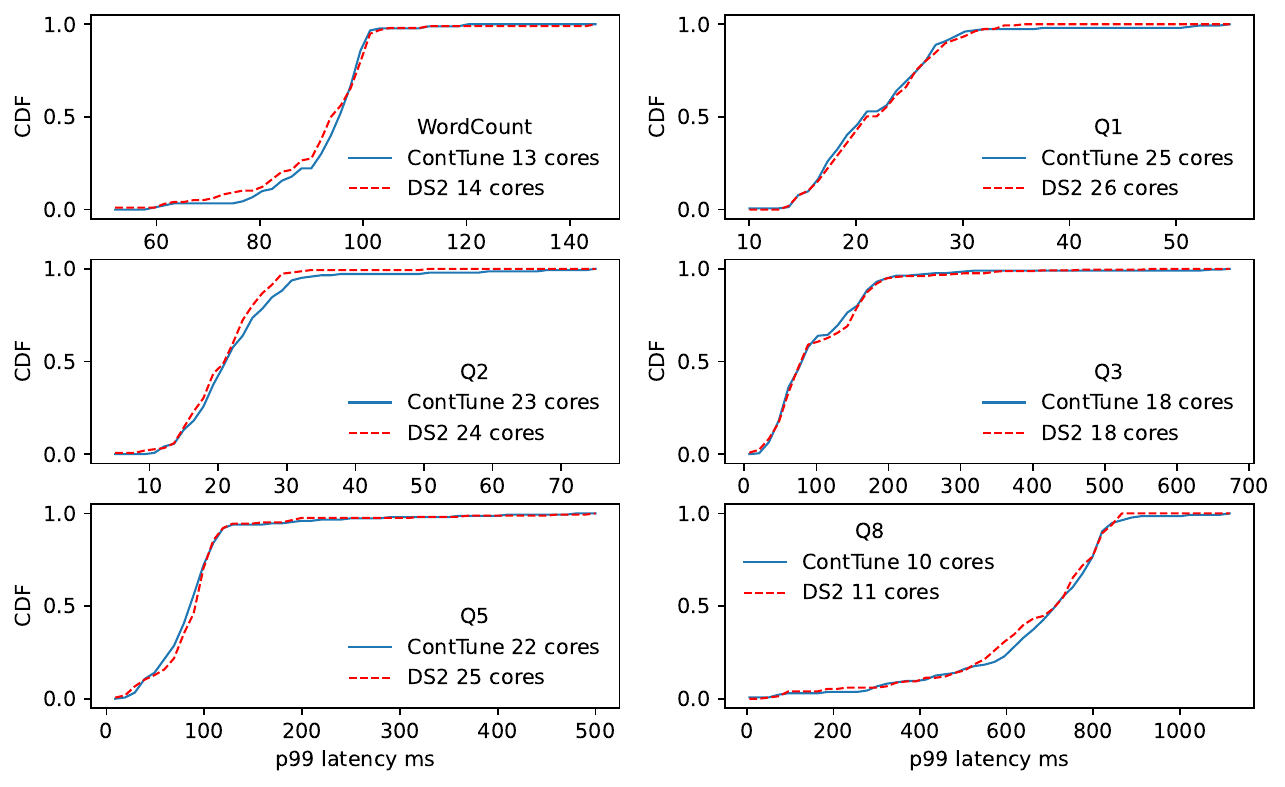}
  \caption{Observed per-record p99 latency CDFs for six quries.}
  \Description{Observed p99 per-record latency CDFs for six quries.}
  \label{Latency}
\end{figure}

\noindent
\textbf{ContTune finds the optimal levels of parallelism via minimal number of reconfigurations.}
Table~\ref{table:ResultsUnderContinuous20SimulatedVariableWorkloads} shows the average number of reconfigurations per tuning to find the optimal levels of parallelism. 
In all cases, ContTune ($\alpha = 3$) takes the minimum number of reconfigurations. 
This shows that ContTune is the most efficient tuning method, and ContTune ($\alpha = 0$) has reduced average \textbf{35.42\%} ($\frac{(2.40-1.55)}{2.40}$) the number of reconfigurations compared to DS2 and ContTune ($\alpha = 3$) has reduced average \textbf{46.25\%} ($\frac{(2.40-1.29)}{2.40}$) the number of reconfigurations compared to DS2. 
For this result, we analyze that it may be brought by \textit{Top-K} and \textit{mean-reversion}, so we make the ablation study of Top-K and mean-reversion on \textbf{Q5} in Section~\ref{sec:experimentAnalysisOfContTune}. 
Due to the ability of the GP to fit the processing ability of the level of parallelism near the observed level of parallelism in historical observations $H^t$, which helps ContTune hit the minimal number of CPU cores. 
In all experiments, ContTune applies the minimal number of CPU cores at the end of each tuning as shown in Table~\ref{table:CpuCoresWhenMaxThroughput}. 
Figure~\ref{Latency} shows the stable job latency of ContTune and DS2 (SOTA method) for the same maximal workload at the end of tuning is similar in 6 jobs, where ContTune applies 1,1,1,0,3,1 (cf. Table~\ref{table:CpuCoresWhenMaxThroughput}) CPU cores less than DS2. 
Despite using fewer CPU resources, the latency of ContTune tuned jobs is essentially the same as that of DS2 tuned jobs. 

\noindent
\textbf{ContTune finds the optimal levels of parallelism via minimal running time.}
The total end-to-end running time as show in Table~\ref{table:E2ERunningTimeResultsUnderContinuous20SimulatedVariableWorkloads} consists of three parts (1) job ideal running time, 72000 (s); (2) time of processing buffered data as show in Table~\ref{table:totalTimeSolvingBufferedData}; (3) tuning time (including the time of reconfigurations) as shown in Table~\ref{table:E2ETuningTimeResultsUnderContinuous20SimulatedVariableWorkloads}. 
Each source of the job generates data for 600 seconds~\footnote{We stop the generation of data when the real data generation arrives at 600 seconds.}, so the job ideal running time is $600 \times 120 \ (20 \times 6 \ pers)   = 72000 \ (s)$.
The inappropriate configurations will make job under-provisioned and unprocessed data buffered in the queue, and needs time to solve these buffered data. 
Besides, each method spends time on finding the optimal levels of parallelism by making reconfigurations. 
In all cases, ContTune achieves both minimum running time in Table~\ref{table:E2ERunningTimeResultsUnderContinuous20SimulatedVariableWorkloads} and tuning time in Table~\ref{table:E2ETuningTimeResultsUnderContinuous20SimulatedVariableWorkloads} compared to other methods. 
This shows that ContTune is the most efficient tuning method, and ContTune ($\alpha = 0$) has reduced average \textbf{2.52\%} ($\frac{(464461.52-452768.28)}{464461.52}$) end-to-end running times compared to DS2 and ContTune ($\alpha = 3$) has reduced average \textbf{3.12\%} ($\frac{(464461.52-449962.35)}{464461.52}$) end-to-end running times compared to DS2. 
ContTune ($\alpha = 0$) has reduced average \textbf{37.36\%} ($\frac{(29877.81-18714.4)}{29877.81}$) end-to-end tuning time compared to DS2 and ContTune ($\alpha = 3$) has reduced average \textbf{47.08\%} ($\frac{(29877.81-15812.5)}{29877.81}$) end-to-end tuning time compared to DS2. 

\begin{table}[t]
\caption{Total time (s) of processing buffered data.}
\label{table:totalTimeSolvingBufferedData}
\scalebox{0.78}{
\begin{tabular}{c|c|c|c|c|c|c}
\hline
Baseline                & WordCount      & Q1              & Q2              & Q3             & Q5              & Q8             \\ \hline
Dhalion                 & 1277.17        & 5966.53         & 2744.8          & 2913.87        & 4644.05         & 513.38         \\ \hline
DS2                     & \textbf{33.66} & 406.35          & \uwave{130.57}          & 193.3          & 1339.06         & 480.77        \\ \hline
Big + DS2               & \uwave{121}            & 334.65          & 148.54          & 192.97         & \uwave{1289.33}         & \textbf{0.04}        \\ \hline
Dragster                & 493.95         & 1006.09         & 724.32          & \textbf{71.24} & \textbf{893.87} & 236.1        \\ \hline
ContTune ($\alpha = 0$) & 123.75         & \uwave{324.45}          & \textbf{112.82} & 159.79         & 1297.18         & \uwave{35.89} \\ \hline
ContTune ($\alpha = 3$) & 265.98         & \textbf{267.18} & 144.47          & \uwave{135.05}         & 1301.28         & \uwave{35.89} \\ \hline
\end{tabular}
}
\vspace{-1em}
\end{table}

\noindent
\textbf{ContTune temporarily requests more CPU Cores and could quickly eliminate the backlogged data.}
We show the maximal number of CPU Cores requested by each method in Table~\ref{table:MaxParallelismResultsUnderContinuous20SimulatedVariableWorkloads} and the total CPU cost (Core $\times$ second) in Table~\ref{table:TotalCostResultsUnderContinuous20SimulatedVariableWorkloads}. 
Dhalion and DS2 request the less maximal number of requested CPU Cores than other methods as shown in Table~\ref{table:MaxParallelismResultsUnderContinuous20SimulatedVariableWorkloads}, because both Dhalion and DS2 tune levels of parallelism from small to big for under-provisioned jobs. 
And the maximum number of CPU Cores requested by ContTune is more than Dhalion and DS2 due to its Big phase. 
However, due to the efficiency of finding the optimal levels of parallelism, the total CPU cost is only a little more than DS2 as shown in Table~\ref{table:TotalCostResultsUnderContinuous20SimulatedVariableWorkloads}. 
In all cases, Dhalion achieves the minimum total CPU cost, ContTune and DS2 use almost the same total CPU cost. 
ContTune ($\alpha = 3$) uses average \textbf{0.22\%} ($\frac{(4376785.59-4367136.52)}{4367136.52}$) total CPU cost smaller than DS2 and ContTune ($\alpha = 3$) uses average \textbf{1.84\%} ($\frac{(4367136.52-4288435.6)}{4288435.6}$) total CPU cost bigger than Dhalion. 
So ContTune temporarily requests more CPU Cores. 

When the job is reconfigured, the data in the processing queue that have not been processed are the backlogged data as shown in Table~\ref{table:BckloggedDataPercentResultsUnderContinuous20SimulatedVariableWorkloads}. 
These data must wait until the job completes the reconfiguration before they can be processed (e.g., Flink, Samza and Heron use the kill-and-restart method to execute reconfigurations~\cite{mao2021trisk}), and the waiting time will increase the job latency. 
Table~\ref{table:BckloggedDataPercentResultsUnderContinuous20SimulatedVariableWorkloads} shows that Big + DS2, ContTune ($\alpha = 0$) and ContTune ($\alpha = 3$) have less backlogged data than DS2, Dragster and Dhalion thanks to the Big phase. 
Both ContTune ($\alpha = 0$) and ContTune ($\alpha = 3$) achieve the best or second-best result as shown in Table~\ref{table:BckloggedDataPercentResultsUnderContinuous20SimulatedVariableWorkloads}, and ContTune ($\alpha = 3$) has reduced average \textbf{89.09\%} ($\frac{(117.33-12.8)}{117.33}$) number of backlogged data compared to Dhalion and ContTune ($\alpha = 3$) has reduced average \textbf{43.01\%} ($\frac{(22.46-12.8)}{22.46}$) number of backlogged data compared to DS2. 
Big + DS2, ContTune ($\alpha = 0$) and ContTune ($\alpha = 3$) also have less time of processing these backlogged data than DS2, Dragster and Dhalion thanks to the Big phase as shown in Table~\ref{table:totalTimeSolvingBufferedData}, and ContTune ($\alpha = 3$) has reduced average \textbf{88.10\%} ($\frac{(18059.8-2149.85)}{18059.8}$) time of processing backlogged data compared to Dhalion and ContTune ($\alpha = 3$) has reduced average \textbf{16.79\%} ($\frac{(2583.71-2149.85)}{2583.71}$) time of processing backlogged data compared to DS2. 
The Big phase temporarily requests more CPU Cores in order to quickly eliminate these backlogged data. 

\begin{table}[t]
\caption{Tuning WordCount on synthetic workloads, $\kappa$ means all the number of reconfigurations, $\theta$ means the number of reconfigurations for eliminating backpressure and $\zeta$ means the number of reconfigurations for over-provisioned jobs.}
\label{FastStability}
\begin{tabularx}{.485\textwidth}{X|X|X|X|X|X|X}
\hline
\multirow{2}{*}{} & \multicolumn{2}{c|}{$\kappa$}              & \multicolumn{2}{c|}{$\theta$}              & \multicolumn{2}{c}{$\zeta$}               \\ \cline{2-7} 
                  & \multicolumn{1}{c|}{DS2} & \multicolumn{1}{c|}{ContTune} & \multicolumn{1}{c|}{DS2} & \multicolumn{1}{c|}{ContTune} & \multicolumn{1}{c|}{DS2} & \multicolumn{1}{c}{ContTune} \\ \hline
$per1$              & \multicolumn{1}{c|}{35}  & \multicolumn{1}{c|}{24}       & \multicolumn{1}{c|}{22}  & \multicolumn{1}{c|}{11}       & \multicolumn{1}{c|}{13}  & \multicolumn{1}{c}{10}       \\ \hline
$per2$              & \multicolumn{1}{c|}{38}  & \multicolumn{1}{c|}{23}       & \multicolumn{1}{c|}{25}  & \multicolumn{1}{c|}{13}       & \multicolumn{1}{c|}{13}  & \multicolumn{1}{c}{9}       \\ \hline
$per3$              & \multicolumn{1}{c|}{35}  & \multicolumn{1}{c|}{22}       & \multicolumn{1}{c|}{26}  & \multicolumn{1}{c|}{13}       & \multicolumn{1}{c|}{9}   & \multicolumn{1}{c}{8}        \\ \hline
\end{tabularx}
\end{table}

\noindent
\textbf{ContTune prioritizes the job SLA.}
Table~\ref{FastStability} presents the total reconfigurations to find the optimal levels of parallelism and the number of reconfigurations used to eliminating backpressure.
We observe that ContTune is faster than DS2 in eliminating the under-provisioned jobs and uses less number of reconfigurations to find the optimal levels of parallelism. 
This could be contributed to the design of the Big-small algorithm.
In the Big phase, ContTune quickly allocates sufficient resources for the under-provisioned jobs.
Then, at the beginning of the Small phase, the SLA of the job is satisfied, and the subsequent tuning is used only to improve the CPU resource utilization. 

\begin{figure}[t]
  \includegraphics[width=0.475\textwidth]{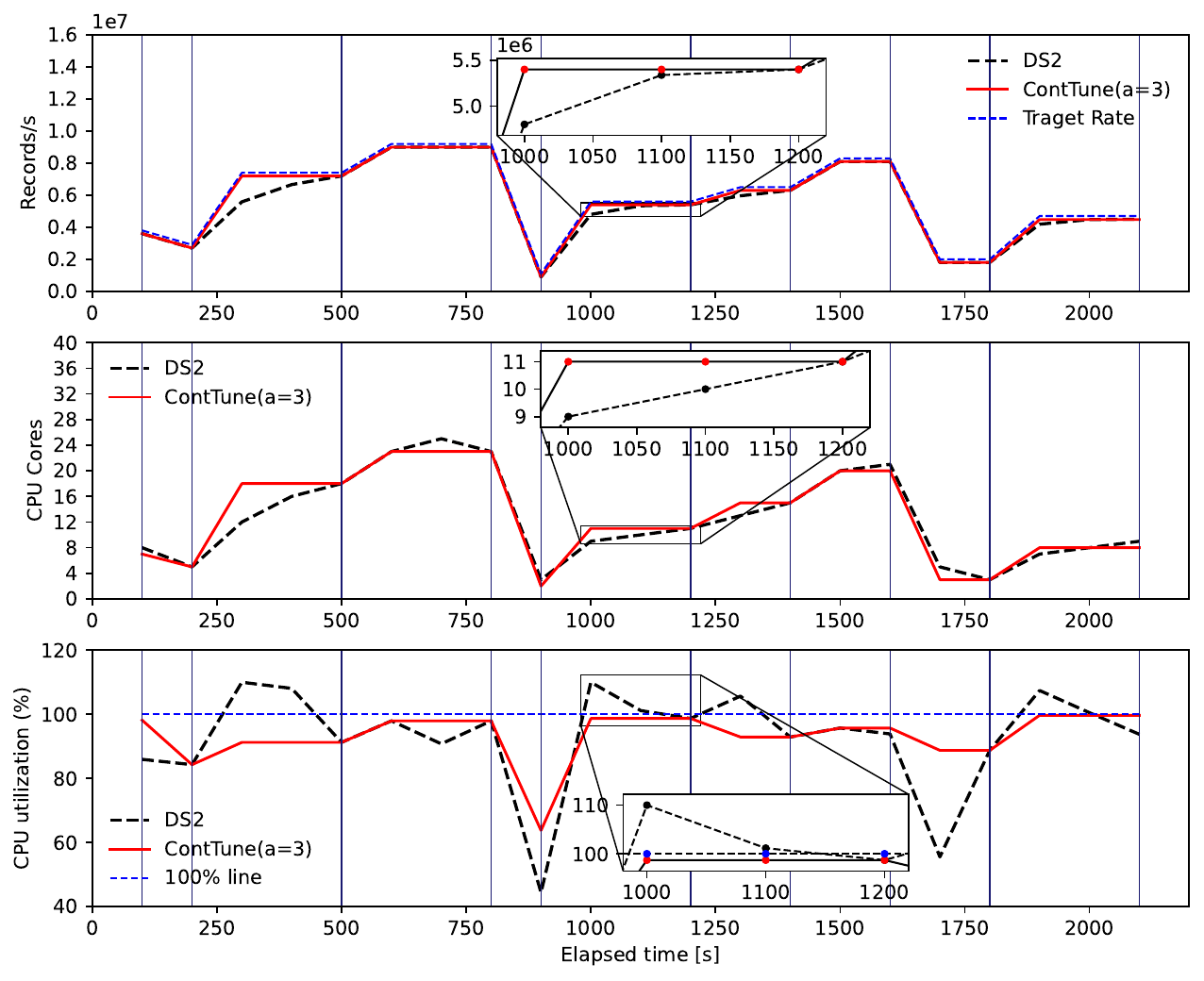}
  \caption{Aggregated Records/s of sources and CPU cores and CPU utilization of Q2 on latter 10 tuning times.}
  \Description{instances of Q2.}
  \label{Query2TenWorkloads}
\end{figure}

\noindent
\textbf{ContTune effectively utilizes the previous tuning observations to speed up the tuning process.}
The Big phase algorithm can quickly eliminate job backpressure and obtain the $\lambda$. 
DS2 employs other methods (e.g., SnailTrail~\cite{DBLP:conf/nsdi/HoffmannLLKDWCR18}) to estimate the $\lambda$. 
Therefore, in this experiment, to validate the efficiency of tuning methods after obtaining the $\lambda$, we proactively obtained the $\lambda$ of each operator, and focused on comparing the efficiency of CBO based on BO with the linear search of DS2 to demonstrate the efficiency of ContTune. 
Figure~\ref{Query2TenWorkloads} presents the performance of latter 10 tuning times for \textbf{Q2}. 
The real CPU utilization does not exceed 100\%. 
Any CPU utilization above the \textit{100\% line} in Figure~\ref{Query2TenWorkloads} means that the job is \textit{under-provisioned}.
We observe that ContTune performs better than DS2 in the latter 10 tuning times.
It takes full advantage of historical observations in the face of a workload that has processed before, rather than starting from the scratch like DS2. 
In all 10 tuning times, the number of reconfigurations used for ContTune is smaller than or equal to the number of reconfigurations used for DS2.
And, the CPU cores used is smaller than or equal to the CPU cores used for DS2, and the CPU cores used at the $1^{st}$, $5^{th}$, $8^{th}$, and $10^{th}$ tuning time smaller than that used for DS2. 

\subsection{Evaluations on Real Workloads}\label{sec:experiment4RealWorkloads}

\begin{figure}[t]
  \includegraphics[width=0.475\textwidth]{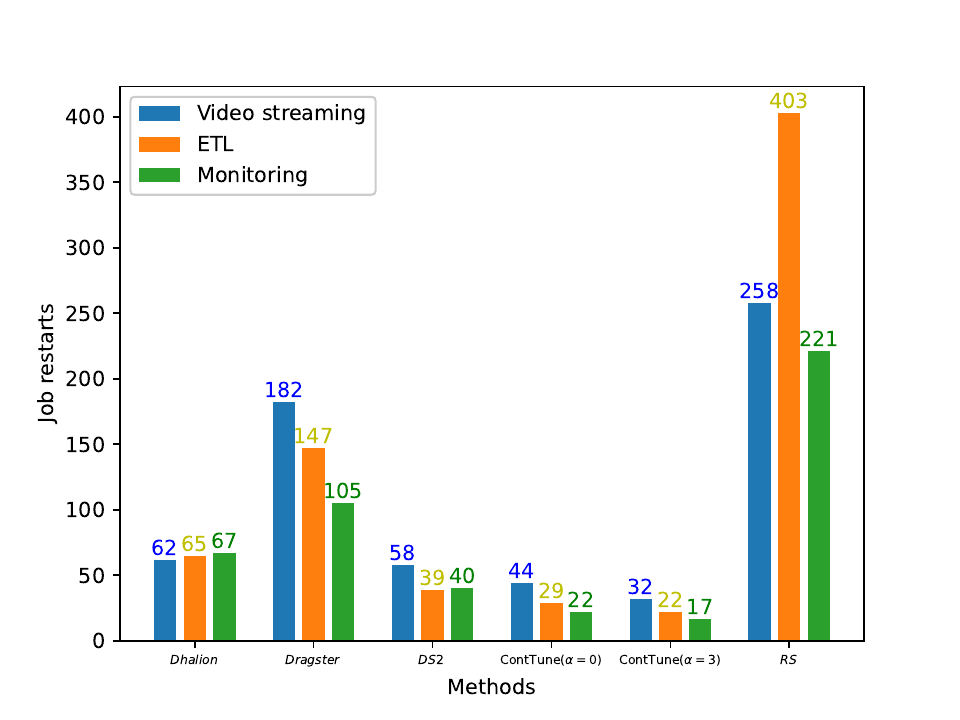}
  \caption{Total reconfigurations on real workloads.}
  \Description{Total reconfigurations on real workloads.}
  \label{realWorkloadsExperiment}
\end{figure}

Figure~\ref{realWorkloadsExperiment} shows the total number of reconfigurations on real workloads. 
Since real workloads do not significantly vary as much as synthetic workloads, the number of tuning may also vary depending on the \textit{controller}, for example, \textit{controller} triggers 26, 21 and 15 tuning times for \textbf{Video Streaming}, \textbf{ETL} and \textbf{Monitoring}. 
Figure~\ref{realWorkloadsExperiment} shows that compared to DS2, ContTune ($\alpha = 0$) reduced \textbf{24.14\%} ($\frac{(58-44)}{58}$) the number of reconfigurations on \textbf{Video streaming}, and \textbf{25.64\%} ($\frac{(39-29)}{39}$) the number of reconfigurations on \textbf{ETL}, and  reduced \textbf{45\%} ($\frac{(40-22)}{40}$) the number of reconfigurations on the \textbf{Monitoring}. 
And ContTune ($\alpha = 3$) reduced \textbf{44.83\%} ($\frac{(58-32)}{58}$) the number of reconfigurations on \textbf{Video streaming}, and \textbf{43.59\%} ($\frac{(39-22)}{39}$) the number of reconfigurations on \textbf{ETL}, and reduced \textbf{57.5\%} ($\frac{(40-17)}{40}$) the number of reconfigurations on \textbf{Monitoring}. 
The main reason that ContTune ($\alpha = 0$) and Dragster are not as efficient as the case on synthetic workloads is that the period of real workload we captured do not necessarily contain multiple workload replication, making it unlikely to apply a simple caching mechanism. 
So the efficiency of the above methods is compromised. 
In ContTune ($\alpha = 3$) the fitting ability of the GP compensates for this drawback better.

\subsection{Analysis of ContTune}\label{sec:experimentAnalysisOfContTune}

\begin{table}[t]
\caption{Total reconfigurations of Q1 on synthetic workloads with different acquisition function (AF).}
\label{DifferentAcquisitionFunctionExperiment}
\begin{tabular}{l|c|c|c|c|c|c|c}
\hline
Baseline & $per1$ & $per2$ & $per3$ & $per4$ & $per5$ & $per6$ & $sum$ \\ \hline
DS2 & 51   & 44   & 47   & 41   & 46   & 46   & 275 \\ \hline
CBO (AF~\ref{EIC}) & 26   & 25   & 27   & 28   & 32   & 26   & 164 \\ \hline
CBO (AF~\ref{equ:acq-our}) & 22   & 23   & 27   & 28   & 25   & 26   & 151 \\ \hline
\end{tabular}
\end{table}

\noindent
\textbf{Comparison of Different Acquisition Functions.}
We propose a carefully designed acquisition function (Equation~\ref{equ:acq-our}) that allows ContTune to suggest the optimal levels of parallelism while strictly satisfying SLA, and we compare it with CEI (Equation~\ref{EIC}). 
Table~\ref{DifferentAcquisitionFunctionExperiment} shows that CBO with Equation~\ref{equ:acq-our} has less number of reconfigurations than CBO with CEI (Equation~\ref{EIC}). 
CEI does not consider the constraint safety-critical, and it may suggest infeasible levels of parallelism during tuning (e.g., trying the level of parallelism $p_i$ with large $ p_i^*-p_i$ but small $Pr[f(p_i)\ge\lambda ]$).
Once these levels of parallelism are suggested, additional reconfigurations are required to keep the job from backpressure. 
Mean represents exploitation in BO, Equation~\ref{equ:acq-our} uses only mean and the surrogate model compose of \textit{fast exploitation}. 
These designs will avoid re-creating application under-provisioned in the Small phase and reduce the number of reconfigurations. 

\begin{table}[t]
\caption{Total reconfigurations of Q5 on three sub-permutations.}
\label{SystemNoisesExperiment}
\begin{tabular}{c|c|c|c}
\hline
Baseline & $per1$ & $per2$ & $per3$ \\ \hline
DS2 & 116   & 66   & 68   \\ \hline
CBO (Without Top-K, mean-reversion) & 50   & 35   & 29   \\ \hline
CBO (With Top-K, mean-reversion)    & 27   & 23   & 26   \\ \hline
\end{tabular}
\end{table}

\noindent
\textbf{Ablation Study of Top-K and Mean-reversion.}
We verified the role of Top-K with mean-reversion by ablation study. 
Since Top-K and mean-reversion can handle noise better and the \textit{window} operator is noisier, we conducted experiments on \textbf{Q5} with mainly the \textit{window} operator. 
Table~\ref{SystemNoisesExperiment} shows CBO (With Top-K, mean-reversion) completes the tuning with fewer number of reconfigurations than CBO (Without Top-K, mean-reversion). 
Top-k and mean-reversion compensate well for the disadvantage that the processing ability of the \textit{window} operator is difficult to count, while without Top-k and mean-reversion, the tuning method depends on the processing ability of the most recent window operator recorded at that time.
The uncertainty of it leads to the uncertainty of the number of reconfigurations required for the tuning method, e.g., for the same job with different permutation, $per1$ achieves 50 of reconfigurations, while $per3$ uses only 29. 
And for $per1$ and $per3$, there is not much difference in their workload variations. 
CBO (With Top-K, mean-reversion) reduced \textbf{46\%} ($\frac{(50-27)}{50}$) the number of reconfigurations on $per1$, and reduced \textbf{34.29\%} ($\frac{(35-23)}{35}$) the number of reconfigurations on \textit{per2}, and reduced \textbf{10.34\%} ($\frac{(29-26)}{29}$) the number of reconfigurations on $per3$, with an average of \textbf{30.21\%}. 

\section{Related Work}

\noindent
\textbf{Configuration of distributed stream data processing systems.} 
Many distributed stream data processing systems have a wide range of configuration parameters, and tuning these parameters can improve performance and reduce resource utilization. 
Operator scaling techniques elastically tunes the amount of each operator's needed resource in order to be suitable for workload variations. 
The user can horizontally or vertically scale operators. 
Horizontally scaling deploys parallel instances of the same operator leveraging \textit{Data Parallelization}, and each instance processes a share of the input stream. 
Vertically scaling focuses on tuning computer resource (e.g., CPU time, instance memory) of the existing instance instead of tuning the level of parallelism. 
In this survey~\cite{cardellini2022runtime}, horizontally scaling is more efficient than vertically scaling, so ContTune focuses on horizontally scaling. 
There are many researches for horizontally scaling. 
~\cite{floratou2017dhalion,ahmad2005distributed,gulisano2012streamcloud,castro2013integrating,gedik2013elastic,heinze2014latency,xu2016stela,pravega2017,yin2022adaptive,wu2022dynamic} are rule-based tuning methods, their effect depends on the setting of rules and thresholds, and they often propose different rules and thresholds for different systems, so the applicability of their methods is poor. 
~\cite{kalavri2018three,mei2020turbine} use the performance relation between workload and operator process ability, but they do not know the non-linear relation between the level of parallelism and process ability, so they use other reconfigurations to tune the levels of parallelism. 
~\cite{fischer2015machines, Liu2022OnlineRO} use Bayesian Optimization to tune the levels of parallelism, and the shorts of aggressive exploration brings many reconfigurations, but the use of historical observations is helpful to establish the surrogate model of the level of parallelism and process ability. 

\noindent
\textbf{Bayesian Optimization.} 
Bayesian Optimization (BO) is a SOTA optimization framework for optimizing of expensive-to-evaluate black-box function.
It has been extensively used in many scenarios, including hyperparameter tuning~\cite{DBLP:conf/nips/BergstraBBK11, DBLP:conf/icdm/WistubaSS15, DBLP:conf/sigmod/LibertyKXRCNDSA20}, experimental design~\cite{DBLP:conf/nips/FosterJBHTRG19} and controller tuning~\cite{DBLP:conf/icra/CalandraSPD14, DBLP:conf/icra/MarcoBHS0ST17, DBLP:conf/cluster/FischerGB15,DBLP:conf/ijcai/FiduciosoCSG019}.
BO uses an acquisition function to suggest the next configuration that trades off exploration (i.e., acquiring new knowledge) and exploitation (i.e., making decisions based on existing knowledge)~\cite{DBLP:conf/nips/KrauseO11}. 
Instead of evaluating the expensive black-box function, the acquisition function relies on a surrogate model that is cheap to compute, and thus can be efficiently minimized in each iteration.
BO has been adopted to configure the parameters of data management systems~\cite{DBLP:journals/pvldb/DuanTB09, DBLP:conf/sigmod/AkenPGZ17, DBLP:conf/sigmod/ZhangWCJT0Z021,DBLP:conf/sigmod/KunjirB20,DBLP:journals/pvldb/CeredaVCD21,DBLP:conf/sigmod/ZhangW0T0022,DBLP:journals/pvldb/ZhangCLWTLC22}. 
However, its favor of exploration causes applying  configurations in  unknown region with potentially bad performance, which is unacceptable for mission-critical applications. 
For online tuning with SLA requirement, we propose the CBO algorithm that utilizes the safe configurations generated from linearity-based methods as conservative exploration.

\section{Conclusion}
 
In this paper, we describe and evaluate \textit{ContTune}, a continuous tuning system for elastic stream processing using the \textit{Big-small} algorithm, the Big phase and the Small phase (CBO). 
ContTune uses 
the Big phase to quickly eliminate job backpressure and buffered data in the queue, and decouple tuning from the topological graph. 
The Big phase can quickly satisfy SLA for under-provisioned jobs, and the Small phase can quickly find optimal the level of parallelism for over-provisioned jobs. 
CBO uses GP as the surrogate model to fit the non-linear relationship for continuous tuning and introduces the SOTA \textit{one-shot parallelism tuning} method as conservative exploration to avoid SLA violations. 
ContTune performs tuning with $O(1)$ average complexity of the number of reconfigurations. 
ContTune achieves the best results for benchmarks or real applications, synthetic or real workloads, compared to the SOTA method DS2.


\bibliographystyle{ACM-Reference-Format}
\bibliography{sample}

\end{document}